\documentclass[twocolumn, trackchanges]{aastex631}
\usepackage{amsmath}
\usepackage{graphicx}
\usepackage{epsf}
\usepackage{color}
\usepackage{enumitem}
\usepackage{csvsimple,booktabs}
\usepackage[flushleft]{threeparttable}
\usepackage{pifont}
\begin{document}
%%%%%%%%%%%%%%%%%%%%%%%%%%%%%%%%%%%%%%%%%%%%%%%%%%%%%%%%%%%%%%%%%%%%%%
\title{\textbf{Efficient Ionizers with Low H$\boldsymbol{\beta}$+[OIII] Equivalent Widths: JADES Spectroscopy of a Peculiar High-z Population}}
%%%%%%%%%%%%%%%%%%%%%%%%%%%%%%%%%%%%%%%%%%%%%%%%%%%%%%%%%%%%%%%%%%%%%%%%%% I will put these in order eventually
%% Isaac L.
\author[0000-0003-4323-0597]{Isaac H. Laseter}
\email{Laseter@wisc.edu}
\affiliation{Department of Astronomy, University of Wisconsin-Madison, Madison, WI 53706, USA}
% Michael M.
\author[0000-0003-0695-4414]{Michael V. Maseda}
\affiliation{Department of Astronomy, University of Wisconsin-Madison, Madison, WI 53706, USA}
% Charlotte S.
\author[0000-0003-4770-7516]{Charlotte Simmonds}
\affiliation{Kavli Institute for Cosmology, University of Cambridge, Madingley Road, Cambridge, CB3 0HA, UK}
\affiliation{Cavendish Laboratory, University of Cambridge, 19 JJ Thomson Avenue, Cambridge, CB3 0HE, UK}
% Ryan E.
\author[0000-0003-4564-2771]{Ryan Endsley}
\affiliation{Department of Astronomy, University of Texas, Austin, TX 78712, USA}
% Dan S.
\author{Daniel Stark}
\affiliation{Steward Observatory, University of Arizona, 933 North Cherry Avenue, Tucson, AZ 85721, USA}
% Andrew B.
\author[0000-0002-8651-9879]{Andrew J. Bunker }
\affiliation{Department of Physics, University of Oxford, Denys Wilkinson Building, Keble Road, Oxford OX1 3RH, UK}
% Rachana B.
\author[0000-0003-0883-2226]{Rachana Bhatawdekar}
\affiliation{European Space Agency (ESA), European Space Astronomy Centre (ESAC), Camino Bajo del Castillo s/n, 28692 Villanueva de la Cañada, Madrid, Spain}
% Kristan B.
\author[0000-0003-4109-304X]{Kristan Boyett}
\affiliation{Department of Physics, University of Oxford, Denys Wilkinson Building, Keble Road, Oxford OX1 3RH, UK}
% Alex C.
\author[0000-0002-0450-7306]{Alex J.\ Cameron}
\affiliation{Department of Physics, University of Oxford, Denys Wilkinson Building, Keble Road, Oxford OX1 3RH, UK}
% Stefano C.
\author[0000-0002-6719-380X]{Stefano Carniani}
\affiliation{Scuola Normale Superiore, Piazza dei Cavalieri 7, I-56126 Pisa, Italy}
% Mirko C.
\author[0000-0002-2678-2560]{Mirko Curti}
\affiliation{European Southern Observatory, Karl-Schwarzschild-Strasse 2, 85748 Garching, Germany}
% Zhiyuan J.
\author[0000-0001-7673-2257]{Zhiyuan Ji}
\affiliation{Steward Observatory, University of Arizona, 933 North Cherry Avenue, Tucson, AZ 85721, USA}
% Pierluigi R.
\author[0000-0002-5104-8245]{Pierluigi Rinaldi}
\affiliation{Steward Observatory, University of Arizona, 933 North Cherry Avenue, Tucson, AZ 85721, USA}
% Aayush S.
\author[0000-0001-5333-9970]{Aayush Saxena}
\affiliation{Department of Physics, University of Oxford, Denys Wilkinson Building, Keble Road, Oxford OX1 3RH, UK}
\affiliation{Department of Physics and Astronomy, University College London, Gower Street, London WC1E 6BT, UK}
% Sandro T.
\author[0000-0002-8224-4505]{Sandro Tacchella}
\affiliation{Kavli Institute for Cosmology, University of Cambridge, Madingley Road, Cambridge, CB3 0HA, UK}
\affiliation{Cavendish Laboratory, University of Cambridge, 19 JJ Thomson Avenue, Cambridge, CB3 0HE, UK}
% Chris W.
\author[0000-0002-4201-7367]{Chris Willott}
\affiliation{NRC Herzberg, 5071 West Saanich Rd, Victoria, BC V9E 2E7, Canada}
% Joris W.
\author[0000-0002-7595-121X]{Joris Witstok}
\affiliation{Cosmic Dawn Center (DAWN), Copenhagen, Denmark}
\affiliation{Niels Bohr Institute, University of Copenhagen, Jagtvej 128, DK-2200, Copenhagen, Denmark}
% Yongda Zhu
\author[0000-0003-3307-7525]{Yongda Zhu}
\affiliation{Steward Observatory, University of Arizona, 933 North Cherry Avenue, Tucson, AZ 85721, USA}
%%%%%%%%%%%%%%%%%%%%%%%%%%%%%%%%%%%%%%%%%%%%%%%%%%%%%%%%%%%%%%%%%%%%%%
\defcitealias{Endsley_2023}{E23}
\defcitealias{Tang_2019}{T19}
\defcitealias{Davis_2023}{D23}
%%%%%%%%%%%%%%%%%%%%%%%%%%%%%%%%%%%%%%%%%%%%%%%%%%%%%%%%%%%%%%%%%%%%%%
\begin{abstract}
    Early JWST photometric studies discovered a population of UV faint ($\rm <L^{*}_{UV}$) $z \sim 6.5-8$ Lyman break galaxies with spectral energy distributions implying young ages ($\sim10$~Myr) yet relatively weak H$\beta$+[OIII] equivalent widths (EW$\rm_{H\beta+[OIII]} \approx 400\AA$). These galaxies seemingly contradict the implicit understanding that young star-forming galaxies are ubiquitously strong H$\beta$+[OIII] emitters, i.e., extreme emission line galaxies (EW $\rm \gtrsim 750\AA$). Low metallicities, high Lyman continuum escape fractions, and rapidly declining star-formation histories have been proposed as primary drivers behind low H$\beta$+[OIII] equivalent widths, but the blend of H$\beta$+[OIII] in photometric studies makes proving one of these scenarios difficult. We aim to characterize this peculiar population with deep spectroscopy from the JWST Advanced Deep Extragalactic Survey (JADES). We find that a significant subset of these galaxies at $z\gtrsim2$ with modest  H$\beta$+[OIII] equivalent widths ($\rm \approx 300\AA-600\AA$) have high ionization efficiencies ($\rm \log \xi_{ion} \gtrsim 25.5~[Hz~erg^{-1}]$). Suppressed [OIII] EW values yet elevated H$\alpha$~and H$\beta$~EW values imply that the level of chemical enrichment is the primary culprit, supported by spectroscopic measurements of metallicities below 12+log(O/H)$\rm \approx 7.70~(10\%Z_{\odot})$. We demonstrate that integrated H$\beta$+[OIII] selections (e.g., H$\beta$+[OIII] EW $> 700$\AA) exclude the most metal-poor efficient ionizers and favor 1) more chemically enriched systems with comparable extreme radiation fields and 2) older starbursting systems. In contrast, metallicity degeneracies are reduced in H$\alpha$ space, enabling the identification of these metal-poor efficient ionizers by their specific star-formation rate.
\end{abstract}
%%%%%%%%%%%%%%%%%%%%%%%%%%%%%%%%%%%%%%%%%%%%%%%%%%%%%%%%%%%%%%%%%%%%%%
\section{\textbf{Introduction}} \label{Introduction}
%%%%%%%%%%%%%%%%%%%%%%%%%%%%%%%%%%%%%%%%%%%%%%%%%%%%%%%%%%%%%%%%%%%%%%
The epoch of reionization, the last major phase transition of the Universe where intergalactic neutral Hydrogen became ionized, likely results from the formation of the first stars and galaxies in the Universe. Considering the pace of Reionization  \cite[$50\%$ completion by $z \sim 7-8$; e.g.,][]{Becker_2015, Planck_2020, Bosman_2022}, rapidly decreasing quasar luminosity functions above $z \gtrsim 3$ \citep[e.g.,][]{Hopkins_2007, Willot_2010, Shen_2020}, and steep faint-end UV continuum luminosity functions ($\alpha \sim 2$) increasing towards higher $z$ \citep[e.g.,][]{Bouwens_2015, Finkelstein_2015, Donnan_2023}, star-forming galaxies (SFGs) have advanced as the primary photon source for Reionization. However, the degree to which these SFGs contribute to Reionization, i.e., the physical mechanisms governing ionizing photon production and escape in the formation and evolution of galaxies, is extensively debated.

The effort in characterizing reionization-era SFGs began with establishing the presence of strong emission lines. Initial stellar population synthesis models applied to multi-band $z \gtrsim 6$ photometry from HST and Spitzer \citep[e.g.,][]{Eyles_2005, Eyles_2007, Bouwens_2010, Gonzalez_2010, Labbe_2010} suggested old ($> 100$~Myr) and massive ($\rm > 10^{9}M_{\odot}$) stellar populations, immediately being at odds with galaxy model predictions of specific star-formation rate (sSFR) evolution \citep[e.g.,][]{Nagamine_2010, Weinmann_2011, Dave_2011}. The crucial caveat with these results was the exclusion of nebular emission and continuum in the population synthesis models. Strong nebular emission can photometrically masquerade as an old population considering the optical side of the Balmer break includes H$\beta$, [OIII]$\lambda\lambda4959,5007$, and H$\alpha$. Studies \citep[e.g.,][]{Schaerer_2009, Ono_2010, Schaerer_2010, Labbe_2013, Smit_2014} demonstrated that incorporating nebular emission lowers SED-derived ages and raises sSFRs, though significant uncertainties in the inferred physical parameters remained considering the nebular contribution and photometric redshifts were derived from the same photometry. \cite{Stark_2013} spectroscopically showed that accounting for nebular contributions decreases SED-derived masses by $\rm \sim 2x$, bringing sSFR measurements in line with theoretical predictions. However, \cite{Stark_2013} could not access $z \gtrsim 5$ since H$\alpha$ becomes unobservable from ground-based telescopes, and weaker lines were impractical to observe. The \textit{James Webb Space Telescope} (JWST) at this time was still approximately a decade away, but progress was still made at $z \lesssim 2$ where the restframe UV/optical was observable. 

\cite{Van_der_well_2011} identified an abundant population of $z \lesssim 2$~extreme emission line galaxies (EELGs) with comparable line strengths to those at higher redshift, but with the added detail that the emission originates from faint UV ($\rm M_{UV} > -20$) and low mass ($\rm \lesssim 10^{8.5}M_{\odot}$) galaxies comprised of young stellar populations ($\rm 1-100$Myr). These results reflected evolutionary models \citep[e.g.,][]{Nagamine_2010, Zheng_2010, Weinmann_2011} exploring intense and episodic star-formation above $z \sim 3$, which was a key differentiating factor of this population relative to local blue-compact dwarfs. For example, the specific star formation rates (sSFR) of EELGs are an order of magnitude higher than local massive SFGs \citep[e.g.,][]{Maseda_2014}. However, so-called ``green peas", discovered in the Sloan Digital Sky Survey (SDSS) around the same time as high-$z$ EELGs, exhibit comparable sSFRs and [OIII]$\lambda 5007$ equivalent widths (EWs) (up to $\rm \sim 1000 \AA$) \citep{Cardamone_2009, Amorin_2010, Jaskot_2013, Henry_2015}. The interpretation of these local studies was that strong emission lines and extremely faint continua originate from low-mass systems undergoing an intense burst of star formation; analogous to the $z \sim 2$~systems. The prevalence of bursty star formation was observed to increase strongly with redshift as gas depletion timescales substantially longer (several hundred Myr) than the mass-weighted ages (50 Myr) \citep[e.g.,][]{Maseda_2018} and nebular line ratios characteristic of high ionization parameters \citep[e.g.,][]{Reddy_2018} were found. Equally important, \cite{Tang_2019} \citepalias{Tang_2019} established EELGs as efficient ionizers using deep MMT and Keck spectroscopy at $z\sim 2$, finding higher Balmer and [OIII]$\lambda\lambda 4959, 5007$ EWs correspond to more efficient ionizing photon production. The overall interpretation emerged that high-$z$ galaxies associated with intense Balmer and [OIII] emission are important contributors to Reionization.

However, JWST calls into question the ubiquity of the youngest galaxies possessing high Balmer and [OIII] EWs. \cite{Endsley_2023} \citepalias{Endsley_2023}, using the Cosmic Evolution Early Release Science Survey (CEERS) and \texttt{Beagle}, found a significant class of sub-$L_{*}$ galaxies with young light-weighted ages ($\lesssim 50$~Myr) yet relatively minor broad-band excesses corresponding to weak H$\beta$ + [OIII] emission ($\rm \lesssim 500\AA$). More evolved systems with weak H$\beta$ + [OIII]$\lambda 5007$ emission are most certainly expected, but a young, starbursting population exhibiting weak H$\beta$ + [OIII] emission would be comparatively rare relative to $z \lesssim 2$~results. \citetalias{Endsley_2023} found the combination of young ages and relatively low H$\beta$ + [OIII] EWs ($\lesssim 600$~\AA) can be reproduced with extremely low metallicities ($\sim 10\%$~$Z_{\odot}$). In such metal-poor models, [OIII] emission is greatly weakened relative to the more ``chemically evolved" galaxies near $\sim 20\%$~$Z_{\odot}$. Even though Balmer EWs increase with decreasing metallicity, the relative contribution of [OIII] to H$\beta$ + [OIII] is generally far more dominant ([OIII]/H$\beta \sim 5-10$; \cite{Steidel_2016, Maseda_2018, Sanders_2018, Tang_2019}) such that extremely low metallicities result in H$\beta$ + [OIII] EWs atypical of $z \lesssim 2$~EELGs. \citetalias{Endsley_2023} was a pure photometric study, so the role of metallicity in driving low H$\beta$ + [OIII] EWs in EELGs is ambiguous as high escape fraction of ionizing photons ($\rm <912\AA;~f_{esc}$) into the intergalactic medium (IGM) and rapidly declining star-formation histories were also proposed.

Equally important, however, it is not clear whether this population is comprised of efficient ionizers. The contribution of ionizing photons to Reionization, i.e. the \textit{ionizing photon budget}, includes the galaxy UV luminosity density ($\rm \rho_{UV}$), the ionizing photon escape fraction ($\rm f_{esc}$), and the ionizing photon production efficiency ($\rm \xi_{ion}$). $\rm \xi_{ion}$ represents the ionizing photon production rate per non-ionizing UV luminosity ($\rm L_{UV}$) around $\rm 1500\AA$:
\begin{equation} \label{Equation 1}
    \rm \xi_{ion}~[Hz~erg^{-1}] = \frac{Q(H_{0})}{L_{UV}},
\end{equation}
with $\rm Q(H_{0})$ representing the intrinsic rate of ionizing photons in units of $\rm s^{-1}$. $\rm \xi_{ion}$~measurements above $z \sim 2$~ have gradually emerged over the last decade, with the primarily photometric results agreeing with the relations from \citetalias{Tang_2019} \citep[e.g.,][]{Stark_2015, Nakajima_2016, Stark_2017, Lam_2019, Maseda2020}. Interestingly, \cite{Emami_2020}, utilizing strong-lensed galaxy clusters (A1689, MACS J0717 and MACS J1149) with Keck/MOSFIRE spectroscopy, found  $1 < z < 3$~low-mass galaxies exhibit higher $\rm \xi_{ion}$ at lower H$\beta$ and [OIII]$\lambda 5007$ EWs compared to the relations from \citetalias{Tang_2019}. In the era of JWST, \cite{Simmonds_2023} determined $\rm \xi_{ion}$ for 30 Ly$\alpha$~emitters in the JWST Extragalactic Medium-band Survey (JEMS, \cite{Williams_2023}), finding Ly$\alpha$~emitters in Reionization have diverse and enhanced $\rm \xi_{ion}$~values ($\rm \log \xi_{ion} \approx 25.5-26.0~[Hz~erg^{-1}]$) compared to literature values \citep{Saxena_2024}. 

These results agree with the first two years of JWST findings that high-$z$~galaxies exhibit large ionization parameters, excitation ratios, nebular EWs, and low metallicities. \citep[e.g.,][]{Cameron_2023, Davis_2023, Katz_2023, Rhoads_2023, Trump_2023}. However, the significant population of young, sub-L$_{*}$ galaxies identified in \citetalias{Endsley_2023} with low H$\beta$ + [OIII]$\lambda 5007$ EWs combined with the elevated ionization efficiencies now routinely measured above $z \sim 2$ suggests there is a population of metal-poor efficient ionizers that are abnormal to the [OIII] trends associated with EELGs identified in \citetalias{Tang_2019}. We aim to spectroscopically characterize the distribution of $\rm \xi_{ion}$ in the high-\textit{z} Universe and attempt to explain the main driver behind low H$\beta$ + [OIII] EWs in young Reionization-era galaxies using medium and deep spectroscopy from the JWST Advanced Deep Extragalactic Survey (JADES).

The remainder of this paper is organized as follows. In Section \ref{Observations} we briefly explain our JADES observations. In Section \ref{Our Sample} we describe our flux and EW measurements and how we subdivide our JADES sample. In Section \ref{The Ionizing Photon Production Efficiency} we define the ionizing photon production efficiency and our methodology. In Section \ref{Metallicity Derivations and Photoionization Modelining} we describe our incorporation of [OIII]$\lambda 4363$~emitters, $\rm \hat{R}$-derived metallicities, and photoionization modeling to further examine trends between nebular EWs and ionizing photon efficiencies. In Section \ref{The Role of Ionization Parameter, Metallicity and Age} we discuss the role of metallicity driving low H$\beta$ + [OIII] EWs. Finally, in Section \ref{Discussion} we discuss the limitations of EELG selections employing [OIII] in identifying extremely metal-poor SFGs. We include a short discussion on Case B departures in the appendix. For this work, we adopt the \cite{Planck_2020} cosmology: H$_0$ = 67.36 km/s/Mpc, $\rm \Omega_{m} = 0.3153$, and $\rm \Omega_{\lambda}$ $= 0.6847$.

%%%%%%%%%%%%%%%%%%%%%%%%%%%%%%%%%%%%%%%%%%%%%%%%%%%%%%%%%%%%%%%%%%%%%%
\section{\textbf{Observations}} \label{Observations}
%%%%%%%%%%%%%%%%%%%%%%%%%%%%%%%%%%%%%%%%%%%%%%%%%%%%%%%%%%%%%%%%%%%%%%

We summarize the JADES NIRSpec and NIRCam observations important for our work below, but we refer the reader to \cite{Bunker_2023},  \cite{Eisenstein_2023}, and \cite{DEugenio_2024} for more general details. Briefly, the JADES program is a joint Guaranteed Time Observation (GTO) program between the NIRSpec and NIRCam Instrument Science Teams comprising two deep pointings and 14 medium-deep pointings in GOODS-S and GOODS-N \citep{Giavalisco_2004} (proposal IDs 1180, 1181, 1210, 1286, and 3215). NIRSpec observations within each visit were performed as a three-shutter nod. The central pointing of each visit was dithered (by $<1$ arcsec) such that common targets were observed in different shutters and detector real estate. Thus, each visit had a unique micro shutter array (MSA) configuration, although the target allocation (performed with the eMPT \footnote{\url{https://github.com/esdc-esac-esa-int/eMPT_v1}; \cite{Bonaventura_2023}}) was optimized for maximizing target commonality between all three dither positions. Priority classes were introduced to maximize NIRSpec's MSA with the highest priority targets \citep{Bunker_2023}. These targets were observed in more than one pointing when possible, so some higher-priority targets have longer integration times. Regardless, the median exposure times across Deep/Medium observations and gratings totaled $\sim 90,000$~s/$\sim 6500$~s (PRISM), $\sim 22,000$~s/$\sim 5400$~s (G140M/F070LP), $\sim 15,000$~s/$\sim 5700$~s (G235M/F170LP), and $\sim 19,400$~s/$\sim 5400$~s (G395/F290LP). Total targets across all pointings number $\sim 4000$  galaxies. For the reduction of NIRCam images, we refer the reader to \cite{Rieke_2023} (JADES DR1), \cite{Eisenstein_2023} (JADES DR2), and \cite{DEugenio_2024} (JADES DR3). These recent photometry releases, used in the current work (Section \ref{UV Luminosities}), cover $\sim 56$ arcmin$\rm ^2$ of NIRCam imaging in GOODS-N and GOODs-S, detecting $\sim 90,000$~distinct objects. These data include 7 overlapping pointings, each with 8-9 separate filters: F090W, F115W, F150W, F182M, F200W, F210M, F277W, F335M, F356W, and F410W. The median exposure time across all filters totaled $\sim 12,000$~s. NIRSpec results are based on version 3 of the JADES reduction pipeline developed by the NIRSpec GTO team and the European Space Agency \citep{Ferruit_2022}. Point source path-loss corrections were considered by modeling each galaxy as a point-like source, considering its relative intra-shutter position \citep{DEugenio_2024}. Extended source path-loss corrections are currently being explored (Curti et al. in prep.).
%%%%%%%%%%%%%%%%%%%%%%%%%%%%%%%%%%%%%%%%%%%%%%%%%%%%%%%%%%%%%%%%%%%%%%
\section{\textbf{Our Sample}} \label{Our Sample}
%%%%%%%%%%%%%%%%%%%%%%%%%%%%%%%%%%%%%%%%%%%%%%%%%%%%%%%%%%%%%%%%%%%%%%
\subsection{\rm \textbf{Emission Line and EW Measurements}} \label{Emission Line and EW Measurements}
%%%%%%%%%%%%%%%%%%%%%%%%%%%%%%%%%%%%%%%%%%%%%%%%%%%%%%%%%%%%%%%%%%%%%%
We are concerned with how $\rm \xi_{ion}$ scales above $z\approx2$ and the dominant mechanism(s) driving low H$\beta$ + [OIII]$\lambda 5007$ EWs in young galaxies observed in \citetalias{Endsley_2023}. Our emission line complexes of interest are therefore [OII]$\lambda\lambda3727, 3729$, H$\gamma$ \& [OIII]$\lambda4363$, H$\beta$ \& [OIII]$\lambda\lambda4959, 5007$, and H$\alpha$ \& [NII]$\lambda\lambda6548, 6583$. We measure fluxes by combining G140M/F070LP, G235M/F170LP, and G395M/F290LP grating/filters and simultaneously fitting Gaussian functions to each emission line complex with free parameters $\lambda$, FWHM, and the respective line amplitudes for $2496$ JADES galaxies. We fit multiple Gaussians simultaneously to increase the fidelity of our fits to weaker emission features. We perform 1000 iterations of our fitting procedure using the observed error-weighted fluxes, taking the median and standard deviation of the fits as our respective fluxes and $68\%$ confidence interval. If [OIII]$\lambda 4959$ or [OIII]$\lambda 5007$ were located within the detector gap we enforced an internal ratio $1/2.98$ of the measured value. We also apply a modest signal-to-noise (S/N) cut of H$\alpha$ \& H$\beta~\geq 5$ and [OIII]$\lambda 5007~\geq 5$ to preserve our initial JADES sample but ensure robust dust corrections and for no conversion from EW(H$\alpha$) to EW(H$\beta$) (i.e., no detector-gap conversion for Balmer lines; see Appendix \ref{Validity of Case B Recombination and the Effects of Dust}); this step narrows our initial sample to $307$~galaxies.

\begin{figure}[hbt!]
    \centering
    \includegraphics[width = \columnwidth]{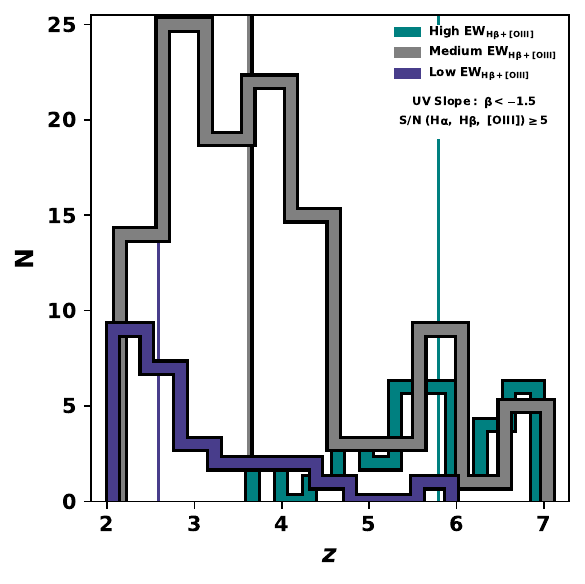}
    \caption{The redshift distribution of our high ($\geq 750$\AA), medium ($200-750$\AA), and low ($\leq 200$\AA) H$\beta$+[OIII] EW samples. We include the median redshift of our respective samples ($\overline{z}_{\rm High}=5.80$, $\overline{z}_{\rm Medium}=3.62$, and $\overline{z}_{\rm Low}=2.59$) and complete sample (black line; $\overline{z}_{\rm Total}=3.69$). The complexity of the JADES selection function is prevalent, and thus we emphasize the current work is not intended as a magnitude-limited population study, but rather an investigation of seemingly atypical systems requiring deep spectroscopy, which JADES thoroughly provides. Our total sample numbers are 31, 116, and 27 for high, medium, and low EW galaxies, respectively.}
    \label{Sample Redshift Distribution}
\end{figure}

It is difficult to apply a single active-galactic-nuclei (AGN) removal for our population as $z \gtrsim 2$ BPT demarcation lines are highly uncertain \citep{Kewley_2013, Maiolino_2023, Maseda_2023, Sanders_2023b}. We therefore visually inspect our line complexes for broadening and match our objects to Chandra observations imposing a $1''$ detection radius; we also crossmatch our sample with AGNs identified in JADES literature \citep[e.g.,][]{Scholtz_2023, Lyu_2024}. Our AGN removal procedure is not unequivocal as high-$z$ AGN are not necessarily detected with Chandra data \citep[e.g.,][]{Maiolino_2024, Pacucci_2024} and broad-line emission is viewing angle dependent, in addition to the peculiar ``little red dot" line widths measured \citep[e.g.,][]{Lambrides_2024}. We emphasize that $\rm \xi_{ion}$ is the ionizing photon efficiency from any ionizing source, but that our steps are necessary for a more pure SFG sample. We further restrict our sample by requiring a UV slope ($\beta$) bluer than $-1.5$~(described in Section \ref{UV Luminosities}) and EW$\rm_{H\alpha} > 100\AA$~as this includes SFGs under a full range of nebular conditions \citep[see][]{Katz_2024}. These steps result in a total of $174$~galaxies.

We now determine the Balmer and [OIII] EWs for the current sample by first carefully inspecting the R100 spectra for clear continua detection. We define a $400$\AA~bin around the emission line complex of interest in the R100 spectrum, removing any emission within this window and taking the median as the continuum level. We then combine our R1000 fluxes and R100 continua to determine observed-frame EWs; we correct for R100 and R1000 flux differences as described in \cite{Bunker_2023}. Our EW fitting procedure is limited as we do not account for stellar absorption in our Balmer fits, though given the strength of the emission lines in our sample the effects of stellar absorption are insignificant \citep{Kong_2002, Groves_2012}. We compare our fitting techniques by employing the penalized pixel fitting algorithm, \texttt{ppxf} \citep{Cappellari_2017, Cappellari_2022}. We omit the full description of the fitting techniques \citep[see][]{Looser_2023} but briefly, \texttt{ppxf} models the continuum as a linear superposition of simple stellar population (SSP) spectra, using non-negative weights and matching the spectral resolution of the observed spectrum. We use the high-resolution (R=10,000) SSP library combining MIST isochrones \citep{Choi_2016} and the C3K theoretical atmospheres \citep{Conroy_2018} within \texttt{ppxf}. We fit for the same emission lines of interest as before using \texttt{ppxf}. We find little deviations in our derived fluxes and EWs between both methods ($\approx 0.08\%$~change). We therefore use the results of our emission line fitting procedure to remain as general as possible, i.e., not appointing a specific suite of SSP models for the analysis.

We correct for dust in our measurements from the available Balmer lines adopting a \cite{Calzetti_2000} attenuation curve. We initially assume the theoretical ratios of H$\alpha / \text{H}\beta = 2.86$, H$\beta / \text{H}\gamma = 2.13$, and H$\alpha / \text{H}\gamma = 6.11$ from Case B recombination at T$=1.0\times 10^4$K. However, $\rm 29\%$ ($\rm 51/176$) of our total sample possesses Balmer ratios less than Case B. These assumed intrinsic ratios are temperature dependent \citep{Osterbrock_2006}, and nebular temperatures are routinely found to be greater than T$=1.0\times 10^4$K \citep[e.g.,][]{Taylor_2022, Laseter_2023, Sanders_2023, Cameron_2024} resulting in a lower intrinsic ratio as the recombination coefficient is dependent on the n-level cross-section and the temperature distribution of electrons. The temperature dependence is typically seen as minor, but the intrinsic ratio of H$\alpha$/H$\beta$ becomes $2.78$ even at T$=1.5\times 10^4$K. We therefore relax our Case B assumption and allow H$\alpha/\text{H}\beta = 2.78$ (corresponding to T$=1.5\times 10^4$K); however, $\rm 26\%$ ($\rm 46/174$) of our total sample still possess Balmer ratios lower than Case B in these conditions. Our findings are not uncommon as Balmer emission lower than Case B and A assumptions are frequent in the high-$z$~Universe \citep[e.g.,][]{Sandles_2023, McClymont_2024, Yanagisawa_2024}. We present a discussion of the effects of dust in Appendix \ref{Validity of Case B Recombination and the Effects of Dust}, but considering we do not have secure electron temperatures for the majority of our sample (Section \ref{[OIII]4363Emitters}), we decide to return to assuming Case B recombination at T$=1.0\times 10^4$K. We do not correct for dust if our measured values are less than these theoretical values, but we identify non-dust-corrected galaxies in the corresponding figures. Ultimately, the self-comparative trends identified in this work are not affected by changing our dust law, but exact $\rm \xi_{ion}~[Hz~erg^{-1}]$~values would likely change due to the respective handling of the UV. Even when employing non-Balmer dust corrections, such as infrared excess $\beta$~relations from \cite{Reddy_2018b}, our results remain.  

%%%%%%%%%%%%%%%%%%%%%%%%%%%%%%%%%%%%%%%%%%%%%%%%%%%%%%%%%%%%%%%%%%%%%%
\subsection{\rm \textbf{UV Luminosities}} \label{UV Luminosities}
\begin{figure*}[hbt!]
    \centering
    \includegraphics[width = \textwidth]{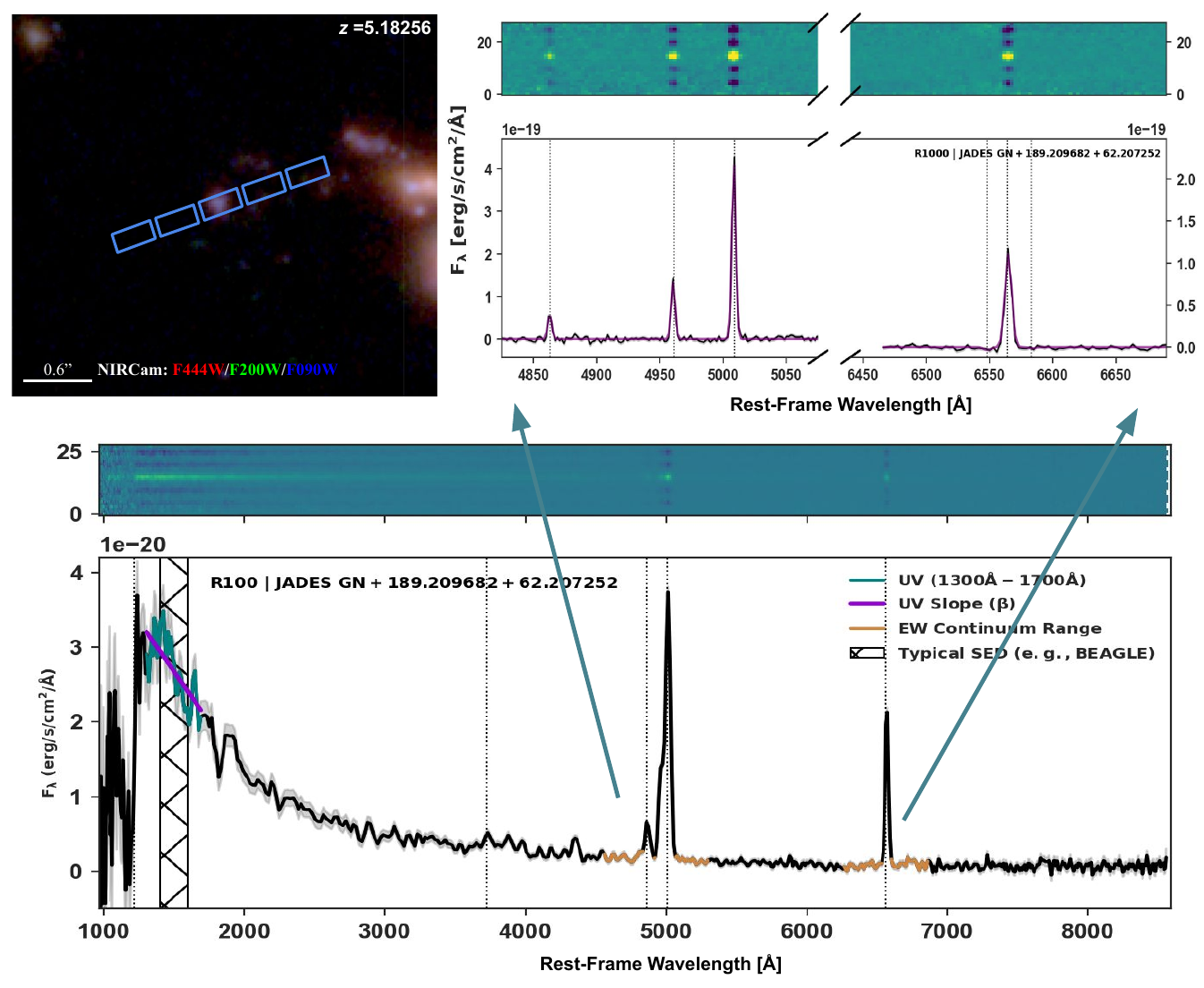}
    \caption{\textit{Top Left}: NIRCam F444W, F200W, and F090W (RGB) image of JADES-GN+189.20968+62.20725 ($z=5.18256$) with NIRSpec MSA shutters overlaid. \textit{Bottom}: R100 spectrum of JADES-GN+189.20968+62.20725. We identify our UV range as green and our derived $\beta$ slope as purple. We shade a $100$\AA~bin centered at $1500$\AA~to demonstrate the common SED-derived UV slopes spectral range. Towards redder wavelengths, we identify the continuum range used in determining Balmer and [OIII] EW as orange ($\rm 200\AA$~around the line center). The top panel shows the 2D spectrum with a clear continuum detection. The lines of interest (located at the vertical dotted lines) shown from left to right are Ly$\rm \alpha$, [OII]$\rm \lambda\lambda3727, 3729$, H$\rm \beta$ \& [OIII]$\rm \lambda\lambda 4959, 5007$, and H$\rm \alpha$. We show our measurement uncertainty as grey-shaded regions. \textit{Top Right}: R1000 spectrum of JADES-GN+189.20968+62.20725. We emphasize our fits for H$\rm \beta$ \& [OIII]$\rm \lambda\lambda 4959, 5007$ and H$\rm \alpha$, which are shown in purple. The top panels show the 2D spectrum for each respective line complex.} 
    \label{Fit Demonstration}
\end{figure*}

%%%%%%%%%%%%%%%%%%%%%%%%%%%%%%%%%%%%%%%%%%%%%%%%%%%%%%%%%%%%%%%%%%%%%%
We first determine the UV luminosities for our full JADES sample following the method outlined in \cite{Hashimoto_2017}. We define a spectral window of $200$\AA~around $1500$\AA, mimicking the typical wavelength range used in spectral energy distribution (SED) codes \citep[e.g.,][]{Chevallard_2016}. We therefore exclude Ly$\alpha$ emission and do not invoke any IGM modeling. We take the definition of UV continuum slopes as $f_{\lambda} \propto \lambda^{\beta}$, so the relation between AB magnitudes \citep{Oke_1983} and our spectral window is:
\begin{equation}
    \rm m_{1500} = -2.5\log([1500\AA\times(1+z)]^{\beta+2}) + A
\end{equation}
where A is the spectral amplitude given in magnitude space. Our absolute magnitude ($\rm M_{UV}$) is therefore given by:
\begin{equation}
    \rm M_{UV} = m_{1500} -5log(d_{L}/10pc) +2.5\log(1+z)
\end{equation}
where $\rm d_{L}$ is the luminosity distance. We determine $\rm M_{UV}$ by applying a Monte Carlo technique for each object, perturbing our spectra by their respective observational errors and fitting the error-weighted spectral window 1000 times. We dust correct our $\rm M_{UV}$ values after converting the nebular color excess found through the Balmer decrement (Section \ref{Emission Line and EW Measurements}) to stellar as described in \cite{Calzetti_2000}, i.e., $\rm E(B-V)_{stellar} = 0.44*E(B-V)_{gas}$. Finally, we determine $\rm L_{UV}$ by taking the median and standard deviation of the total runs as our respective values and $68\%$ confidence interval. We routinely observe CIV$\lambda\lambda 1548, 1551$ emission in our spectra, implying a metal-poor stellar population \citep[e.g.,][]{Topping_2024}. We verify these lines do not affect our $\rm M_{UV}$ measurements by masking  CIV$\lambda\lambda 1548, 1551$ and repeating our procedure, finding no change. We also perform a similar photometric analysis using NIRCam filters F070W, F090W, F115W, F150W, and F200W. We perform the same procedure described above but with a spectral window of approximately $\rm 1300\AA - 2800\AA$, similar to past works \citep[e.g.,][]{Stanway_2005, Bouwens_2009, Wilkins_2011, Hashimoto_2017}. The results of the current analysis remain unchanged when using NIRCam photometry. We demonstrate our spectral UV slope fitting procedure in Figure \ref{Fit Demonstration} using object JADES-GN+189.20968+62.20725. We shade a $200$\AA~bin centered at $1500$\AA~to demonstrate the common SED-derived UV slopes spectral range; we also shade a $200$\AA~EW continuum window centered at the H$\beta$+[OIII] and H$\alpha$~complexes and include R1000 fits from Section \ref{Emission Line and EW Measurements}. 

%%%%%%%%%%%%%%%%%%%%%%%%%%%%%%%%%%%%%%%%%%%%%%%%%%%%%%%%%%%%%%%%%%%%%%
 \subsection{\rm \textbf{The Final Samples}} \label{The Final Samples}
%%%%%%%%%%%%%%%%%%%%%%%%%%%%%%%%%%%%%%%%%%%%%%%%%%%%%%%%%%%%%%%%%%%%%%
The inability to separate H$\beta$ and [OIII] creates difficulties in interpreting photometric results. With NIRSpec MSA spectroscopy we can separate these components to examine the respective nebular contributions and thus the underlying physical mechanisms, as well as the bias introduced when selecting on H$\beta +$[OIII] EW in the high-$z$ Universe (Section \ref{Selection of EELGs/Efficient Ionizers}). We subdivide our sample into respective low, medium, and high H$\beta + $[OIII]~ EW samples; we choose the low H$\beta+$[OIII] EW sample to be $\leq 200$\AA, the medium H$\beta + $[OIII] EW  sample to be between $200-750$\AA, and the high H$\beta + $[OIII] EW sample to be $\geq 750$\AA, reflecting observations of high-$z$ studies \citep[e.g.,][]{Boyett_2022, Davis_2023, Endsley_2023, Boyett_2024}. We summarize our sample in Figure \ref{Sample Redshift Distribution}; $27~(15\%)$ galaxies are within the low EW bin, $116~(67\%)$ within medium, and $31~(18\%)$ within high. We stress, however, that JADES is not a magnitude-limited survey. As such, the current work is not intended as a full population analysis, but rather an investigation of seemingly atypical systems requiring deep spectroscopy, which JADES provides.
%%%%%%%%%%%%%%%%%%%%%%%%%%%%%%%%%%%%%%%%%%%%%%%%%%%%%%%%%%%%%%%%%%%%%%
\section{\rm \textbf{The Ionizing Photon Production Efficiency}} \label{The Ionizing Photon Production Efficiency}
%%%%%%%%%%%%%%%%%%%%%%%%%%%%%%%%%%%%%%%%%%%%%%%%%%%%%%%%%%%%%%%%%%%%%%

\begin{figure*}[hbt!]
    \centering
    \includegraphics[width = \textwidth]{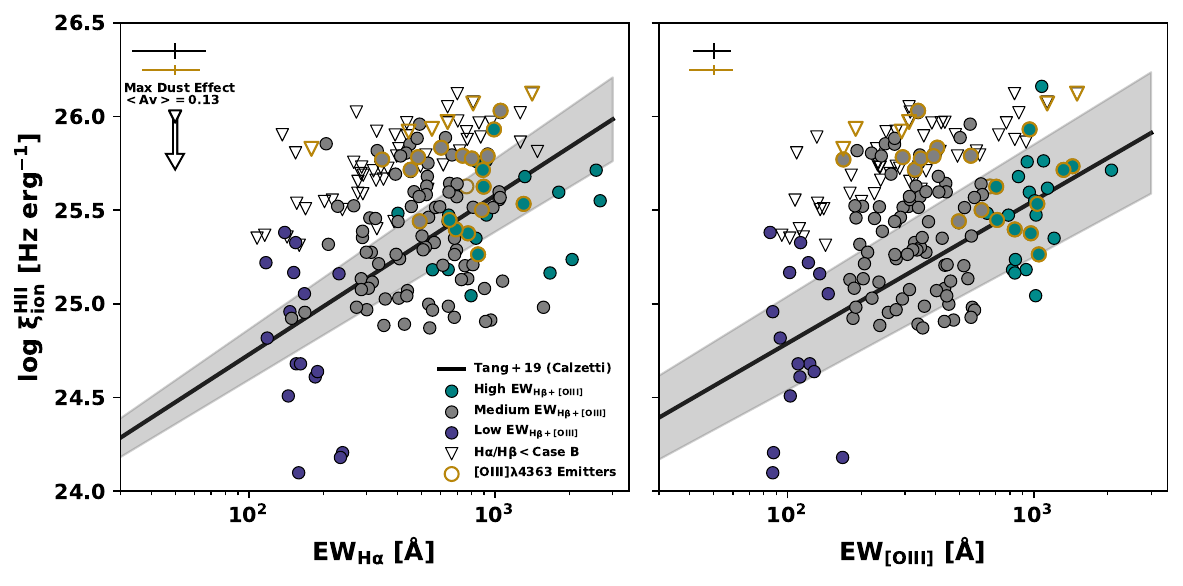}
    \caption{\textit{Left}: Correlation between $\rm \log \xi^{HII}_{ion}$ and H$\alpha$~EW for our high, medium, and low EW samples. We include the HST-derived $z \sim 2$ relation with $68\%$~confidence intervals from \citetalias{Tang_2019}. We largely agree with the relation found from \citetalias{Tang_2019}, but there is a clear overlap between our medium and high EW sample at higher H$\alpha$~EWs. We identify our Case B departure systems as triangles and note the maximum dust effect taken from the average of our low, medium, and high EW samples, though this effect is likely overestimated for this population (see Appendix \ref{Validity of Case B Recombination and the Effects of Dust}). We identify [OIII]$\lambda 4363$~emitters as gold outlines. \textit{Right}: Correlation between $\rm \log \xi^{HII}_{ion}$ and [OIII]$\lambda 5007$~EW for our high, medium, and low EW samples. [OIII] dominance in H$\beta$+[OIII] EWs is evident as there is little overlap in [OIII] space. However, a significant population of our Medium EW sample has $\rm \xi^{HII}_{ion}$~values similar to our high EW sample, suggesting oxygen composition differences at comparable sSFRs. This systematic offset between H$\alpha$~and [OIII] is further evident in deviations from the HST-derived relation from \citetalias{Tang_2019} for [OIII] EW.}
    \label{Ionization Efficiency vs EWs}
\end{figure*}

Several variants of $\rm \xi_{ion}$ exist in the literature, so we explicitly define $\rm \xi_{ion}$ in Equation \ref{Equation 1} as $\rm \xi^{HII}_{ion}$ following the definition from \cite{Chevallard_2018}. $\rm \xi^{HII}_{ion}$ includes the UV luminosity corrected for dust attenuation outside HII regions but not for dust attenuation inside H II regions nor for nebular continuum emission. SED-derived studies typically employ $\rm \xi^{*}_{ion}$, in which $\rm L_{UV}$ is the intrinsic stellar monochromatic UV luminosity ($\rm L^{*}_{UV}$), thus excluding dust and nebular emission/absorption. There is also simply $\rm \xi_{ion}$, which includes the observed (not dust corrected) $\rm L_{UV}$. The various definitions affect quoted results, so we compare against literature $\rm \xi^{HII}_{ion}$ values unless otherwise noted.

$\rm Q(H_0)$ in Equation \ref{Equation 1} is not an observed quantity as there is photon escape from these regions, i.e., $\rm Q_{obs}(H_0) = (1-f_{esc})Q(H_0)$. $f_{esc}$ is largely unconstrained \citep{Mitra_2013, Paardekooper_2015}, resulting in some studies grouping $\rm \xi^{HII}_{ion}$ and $f_{esc}$ as a single parameter \citep[e.g.,][]{Duncan_2015}. There is the added complexity of dust absorption of ionizing photons \citep[e.g.,][]{Tacchella_2022}, which if accounted for increases $\rm Q(H_0)$---the opposite effect of $f_{esc}$. These competing effects are extraneous for our self-comparative work, so we set $f_{esc} = 0$~and apply no correction for $\rm Q(H_0)$~dust absorption. Regardless of what we set these values to, $\rm Q_{obs}(H_0)$ can be converted to an H$\alpha$ luminosity\footnote{Temperature and density will also affect the conversion from $\rm Q(H_0)$~to H$\alpha$, though at the typical temperatures we measure in Section \ref{[OIII]4363Emitters}, these differences are below the systematic trends with metallicity that we identify in this work.} as in \cite{Murphy_2011}:
\begin{equation} \label{Equation 4}
    \rm \xi^{HII}_{ion}~[Hz~erg^{-1}] = 7.38\times10^{11}~(\frac{L_{H\alpha}}{L_{UV}}).
\end{equation}
We use Equation \ref{Equation 4} to determine $\rm \xi^{HII}_{ion}$ for our low, medium, and high H$\rm \beta + [OIII]\lambda 5007$ EW samples. We specifically report the median $\rm \log \xi^{HII}_{ion}$~value for each galaxy after 1000 iterations of perturbing the measurement with our L$\rm_{H\alpha}$~and L$\rm_{UV}$~derived errors; our reported errors are taken as the standard deviation of the distribution of $\rm \xi^{HII}_{ion}$~values. We find the median of our samples to be $\rm \log \xi^{HII}_{ion} = 25.22~[Hz~erg^{-1}]$~(low), $\rm \log \xi^{HII}_{ion} = 25.52~[Hz~erg^{-1}]$~(medium), and $\rm \log \xi^{HII}_{ion} = 25.52~[Hz~erg^{-1}]$~(high) with standard deviations of $0.48$, $0.33$, and $0.27$~dex. We present in Figure \ref{Ionization Efficiency vs EWs} our H$\alpha$ EWs and [OIII]$\lambda 5007$ EWs with $\rm \xi^{HII}_{ion}$, respectively, and compare against HST-derived $z \sim 2$~relations from \citetalias{Tang_2019}. 

Interestingly, we observe a systematic offset between EW$\rm_{H\alpha}$ and EW$\rm_{[OIII]}$ at fixed $\rm \xi^{HII}_{ion}$~between our medium and high EW samples. Specifically, above $\rm \log \xi^{HII}_{ion} \geq 25.5~[Hz~erg^{-1}]$ our medium EW sample exhibits H$\alpha$~EWs comparable to our high EW sample ($\rm \gtrsim 600 \AA$) while simultaneously exhibiting low [OIII] EWs ($\rm \sim 300\AA$). We perform a Welch's t-test (unequal variances t-test) \citep{Welch_1947} between our medium and high EW sample for galaxies with $\rm \log \xi^{HII}_{ion} \geq 25.5~[Hz~erg^{-1}]$. For EW$\rm_{H\alpha}$, we find a t-statistic of $1.17$ with a p-value of $0.25$, indicating that our medium and high EW samples overlap and their respective separation is not statistically significant. In comparison, for EW$\rm_{[OIII]}$, we find a t-statistic of $12.22$ with a p-value of $9.51\times10^{-14}$, indicating the respective samples have different means with strong statistical significance. This distinction in nebular EW and $\rm \xi^{HII}_{ion}$ space is not seen in the $z \lesssim 2$ \citetalias{Tang_2019} sample, indicating a physically motivated difference above $z \gtrsim 2$. Interestingly, \cite{Zhu_2024}, using the Systematic Mid-infrared Instrument Legacy Extragalactic Survey (SMILES) program, found the relation between $\rm \xi^{HII}_{ion}$~and EW$\rm_{[OIII]}$~to be shallower relative to \citetalias{Tang_2019} while the H$\alpha$~relation still agrees, similar to our findings. The selection functions of SMILES \citep{Alberts_2024} and JADES \citep{Bunker_2023, Eisenstein_2023, DEugenio_2024} differ, so it is promising that there are corroborating results, though the underlying causes still have not been explored. Further, \cite{Rinaldi_2023, Rinaldi_2024}, using medium and broadband NIRCam/MIRI imaging, demonstrated diverse [OIII]$\lambda5007$/H$\beta$~ratios ($\rm -1\lesssim log([OIII]\lambda5007/H\beta) \lesssim 1.5$) at high H$\alpha$~EWs and $\rm log\xi^{HII}_{ion}$~values ($\rm \gtrsim 25.5~[Hz~erg^{-1}]$), suggesting the H$\beta$+[OIII] photometric flux excess is not necessarily always driven by [OIII]$\lambda\lambda 4959,5007$~and that elevated $\rm log\xi^{HII}_{ion}$~values are still expected at lower [OIII] EWs. As mentioned, \citetalias{Endsley_2023} proposed extremely low metallicities, high $f_{esc}$, and rapidly declining SFRs as possible solutions to their young, low H$\rm \beta + [OIII]\lambda 5007$ EW sample.

It is difficult for us to comment directly on $f_{esc}$ as we cannot measure it directly, but as $\rm \xi^{HII}_{ion}$~is defined by $\rm Q(H_{0})$, we do not expect a decrease in [OIII] EW at fixed $\rm \xi^{HII}_{ion}$~(the leftward-offset seen in Figure \ref{Ionization Efficiency vs EWs}) if $f_{esc}$ was high. Additionally, the decrease in [OIII] EW at fixed $\rm \xi^{HII}_{ion}$~is apparent for $\rm \log \xi^{HII}_{ion} \gtrsim 25.5~[Hz~erg^{-1}]$, requiring the presence of recently formed O-type stars \citep[][see Section \ref{Selection of EELGs/Efficient Ionizers}]{Maseda2020, Katz_2024}, therefore indicative of recent or ongoing star-formation rather than a decline. Moreover, as we present in the next section, the electron temperatures of galaxies aligning with these trends range between $\rm 10,000-25,000K$, indicative of recent star formation. Extremely low metallicities remain the most likely candidate driving low H$\beta$+[OIII] EWs seen in the current work. We now focus on examining the role of metallicity in $\rm \xi^{HII}_{ion}$ and nebular EWs.
  
%%%%%%%%%%%%%%%%%%%%%%%%%%%%%%%%%%%%%%%%%%%%%%%%%%%%%%%%%%%%%%%%%%%%%%
\section{\textbf{Metallicity Derivations and Photoionization Modeling}} \label{Metallicity Derivations and Photoionization Modelining}
%%%%%%%%%%%%%%%%%%%%%%%%%%%%%%%%%%%%%%%%%%%%%%%%%%%%%%%%%%%%%%%%%%%%%%
\cite{Laseter_2023} and \cite{Sanders_2023} demonstrated a clear failure of locally-derived strong-line calibrations in the high-$z$ Universe due to elevated excitation ratios at fixed metallicity, especially at metallicities lower than $\rm 10\%Z_{\odot}~(12+\log(O/H) \approx 7.7)$. \cite{Sanders_2023} provided high-$z$ calibrations using early CEERS observations, whereas \cite{Laseter_2023} proposed the so-called $\rm \hat{R}$~calibration involving a different projection in the space defined by log([OII]$\lambda 3727,29$/H$\beta$) (R2), log([OIII]$\lambda 5007$/H$\beta$) (R3), and metallicity. We continue with approximating metallicities using the $\rm \hat{R}$~calibration from \cite{Laseter_2023} and the R3 calibration from \cite{Sanders_2023} depending on the S/N of [OII]$\lambda 3727,29$~as described in Section \ref{The Role of Ionization Parameter, Metallicity and Age}. Our following conclusions are unaffected by interchanging these calibrations, but galaxy-by-galaxy cases do differ. We, therefore, use the diagnostic power of JWST to further identify a large sample of [OIII]$\lambda 4363$ emitters to more robustly investigate metallicity in nebular EW and $\rm \xi^{HII}_{ion}$ space.

%%%%%%%%%%%%%%%%%%%%%%%%%%%%%%%%%%%%%%%%%%%%%%%%%%%%%%%%%%%%%%%%%%%%%%
\subsection{\rm \textbf{[OIII]$\lambda 4363$ Emitters}} \label{[OIII]4363Emitters}
%%%%%%%%%%%%%%%%%%%%%%%%%%%%%%%%%%%%%%%%%%%%%%%%%%%%%%%%%%%%%%%%%%%%%%

\begin{figure*}[hbt!]
    \centering
    \includegraphics[width = \textwidth, height=5cm]{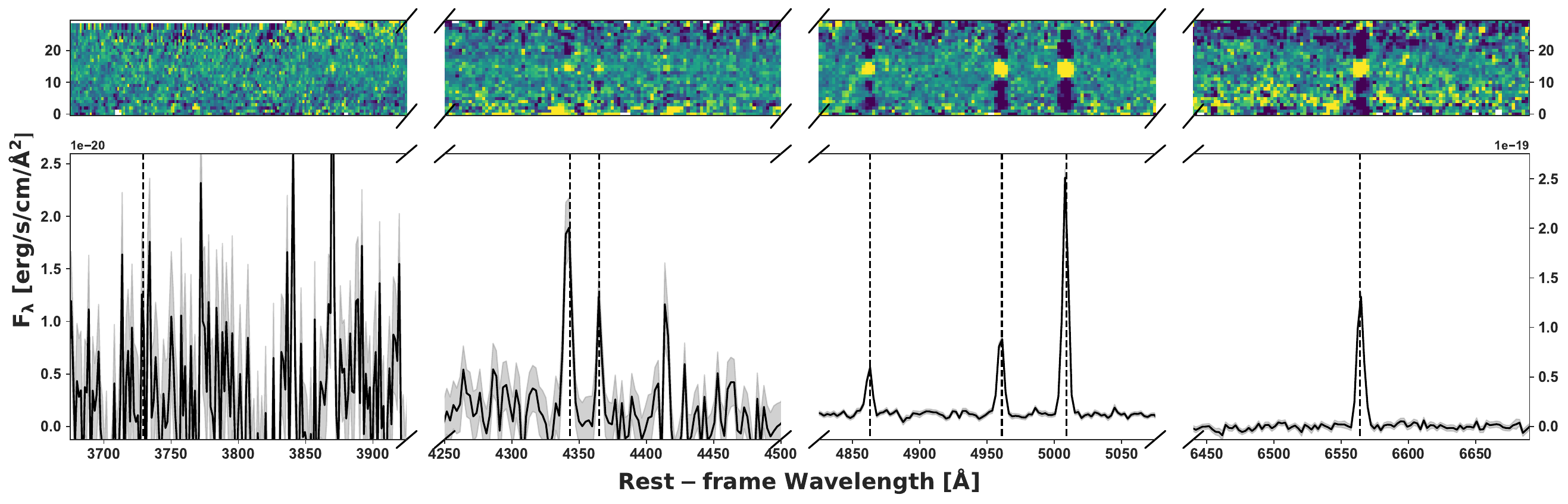}
    \caption{JWST/NIRSpec R1000 spectrum for our newly reported [OIII]$\lambda 4363$~emitter JADES-GN+189.16215+62.26381 with a metallicity of 12+log(O/H)$\rm =7.30 \pm 0.12$~at $z = 6.3$. The lines of interest (located at the vertical dotted lines) shown from left to right are [OII]$\lambda\lambda 3727,3729$, H$\gamma$, [OIII]$\lambda4363$, H$\beta$, [OIII]$\lambda\lambda 4959,5007$, and H$\alpha$. [OII]$\lambda\lambda 3727, 3729$, H$\gamma$ and [OIII]$\lambda 4363$~(H$\beta$, [OIII]$\lambda\lambda 4959,5007$~and $\rm H\alpha$) share the same y-axis. The top panels show the 2D spectrum for each respective line complex. We show our measurement uncertainty as grey-shaded regions.}
    \label{OIIIt spectra}
\end{figure*}

\cite{Laseter_2023} identified $10$ [OIII]$\lambda 4363$ emitters in JADES DEEP-HST \citep{Bunker_2023b} in the Great Observatories Origins Deep Survey South (GOODS-S) legacy field; we have now increased this number to $60$ with our additional JADES observations (Section \ref{Observations}). We follow our analysis as described in Sections \ref{Our Sample} and \ref{The Ionizing Photon Production Efficiency} for this specific sample while deriving electron temperatures and metallcities as described in \cite{Laseter_2023}. Our final [OIII]$\lambda 4363$ sample for the current work is $31 $~ as we impose a [OIII]$\lambda 4363$ S/N restriction $>3$ (the median S/N in our [OIII]$\lambda 4363$ sample is $\sim 4.25$), exclude [OIII]$\lambda 4363$ emitters lacking UV coverage, and remove those with broad Balmer emission and electron temperatures above $\rm T_{e} \geq 30,000$, since these are beyond the physical range associated with the earliest O-type star \citep[e.g., O3, $\rm T_{eff}\approx 40,000K$,][]{Massey_2004}. The latter sample is intriguing based on predictions of hotter stars and nebular continuum contribution \citep{Cameron_2024,Katz_2024}, but these galaxies require more careful electron temperature derivations that we leave for forthcoming work. We present in Figure \ref{OIIIt spectra} the [OII]$\lambda\lambda 3727, 3729$, H$\gamma$ and [OIII]$\lambda 4363$, and H$\beta$, [OIII]$\lambda\lambda 4959, 5007$ complexes for JADES-GN+189.16215+62.26381, a $z = 6.3$ novel [OIII]$\lambda 4363$~emitter with 12+log(O/H)$=7.30\pm0.12$~and $\rm log \xi^{HII}_{ion}=25.8\pm0.01~[Hz~erg^{-1}]$.

In Figure \ref{Ionization Efficiency vs EWs} we include our [OIII]$\lambda 4363$ emitters. Qualitatively, it is clear our [OIII]$\lambda 4363$ sample is representative of the JADES medium and high EW samples in nebular EW and $\rm \xi^{HII}_{ion}$ parameter space, including the leftward systematic offset from the \citetalias{Tang_2019}~[OIII] EW relation apparent in our medium EW sample. Similar to our statistical separation between our medium and high EW samples in [OIII] space, when we compare the statistical separation of our [OIII]$\lambda 4363$ sample to our high EW sample, we find a t-statistic of $3.43$ with a p-value of $0.001$, indicating a statistically significant separation. In contrast, in H$\alpha$~space, we find a t-statistic of $0.01$ with a p-value of $0.99$, indicating no significant separation. We measure t$_{3}$~values between $\rm \approx 13,000K-23,000K$, which are nebular temperatures associated with the presence of OB-type stars and recent star formation, aligning with the notion that the leftward systematic offset at fixed $\rm \xi^{HII}_{ion}$~with EW$\rm_{[OIII]}$ is not from the reduction or loss of ionizing photons. We now incorporate our [OIII]$\lambda 4363$ sample and strong-line derived metallicities with spectral synthesis and photoionization models to further explore the disparities between Balmer and [OIII] EWs with $\rm \xi^{HII}_{ion}$. 

%%%%%%%%%%%%%%%%%%%%%%%%%%%%%%%%%%%%%%%%%%%%%%%%%%%%%%%%%%%%%%%%%%%%%%
\subsection{\rm \textbf{BPASS+CLOUDY}} \label{BPASS+CLOUDY}
%%%%%%%%%%%%%%%%%%%%%%%%%%%%%%%%%%%%%%%%%%%%%%%%%%%%%%%%%%%%%%%%%%%%%%
% In this subsection, we aim to create nebular models to further investigate the dissimilar trends of Balmer and [OIII] EWs with $\rm \log \xi^{HII}_{ion}$. 
The entirety of our derived values depend on the recombination or collisional excitation within HII regions, and thus the nature of the young stellar populations creating these regions. We therefore utilize \texttt{BPASS v2.2.1} \citep{Eldridge_2009} spectral synthesis predictions as our spectral model inputs for our current study. \texttt{BPASS} does not explicitly include nebular emission, so we utilize available \texttt{CLOUDY} \citep{Ferland_2017} processed \texttt{BPASS} models \footnote{Details of these \texttt{CLOUDY} models can be found here: \url{https://bpass.auckland.ac.nz/4.html}.} \citep{Xiao_2018}. In general, these models assume $\rm n_e = 200 cm^{-3}$ while varying $\rm \log(U)$ between $\rm -1.0~and-4.0$ with dust grains included, using the favored model from the \texttt{BPASS} collaboration (\texttt{v2.2.1 imf135\_300}) including a broken upper IMF slope of -1.35, an upper mass limit of 300 $\rm M_{\odot}$, and an instantaneous burst; the covering fraction was set to $1.0$, thus matching our assumption of $f_{esc} = 0$ (Section \ref{The Ionizing Photon Production Efficiency}). Modeling nebular gas conditions with \texttt{CLOUDY} involves an array of free parameters, and thus results can vary widely when comparing individual HII regions to assumed models. We stress that we are not modeling the individual galaxies in our sample, but rather examining global properties driving the systematic offset in Figure \ref{Ionization Efficiency vs EWs}, and thus these \texttt{BPASS+CLOUDY} models suffice for our purposes. We include $\rm \log(U)$ models of $\rm -1.0, -2.0$, and $-3.0$ ranging between $\sim 2$~Myr and $\sim 10$~Myr after the star-formation event. We note for this exercise that we report the model ionization efficiency defined as $\rm \xi^{\star}_{ion}$~in Section \ref{The Ionizing Photon Production Efficiency}, as we are concerned with the intrinsic stellar populations (see Section \ref{Selection of EELGs/Efficient Ionizers} for a discussion of the relation with $\rm \xi^{HII}_{ion}$). We now examine the effects of age, ionization, and metallicity in EW and $\rm \xi^{HII}_{ion}$ space.

%%%%%%%%%%%%%%%%%%%%%%%%%%%%%%%%%%%%%%%%%%%%%%%%%%%%%%%%%%%%%%%%%%%%%%
\section{\rm \textbf{The Role of Ionization Parameter, Metallicity \& Age}} \label{The Role of Ionization Parameter, Metallicity and Age}
%%%%%%%%%%%%%%%%%%%%%%%%%%%%%%%%%%%%%%%%%%%%%%%%%%%%%%%%%%%%%%%%%%%%%%

\begin{figure*}[hbt!]
    \centering
     \includegraphics[width=\textwidth]{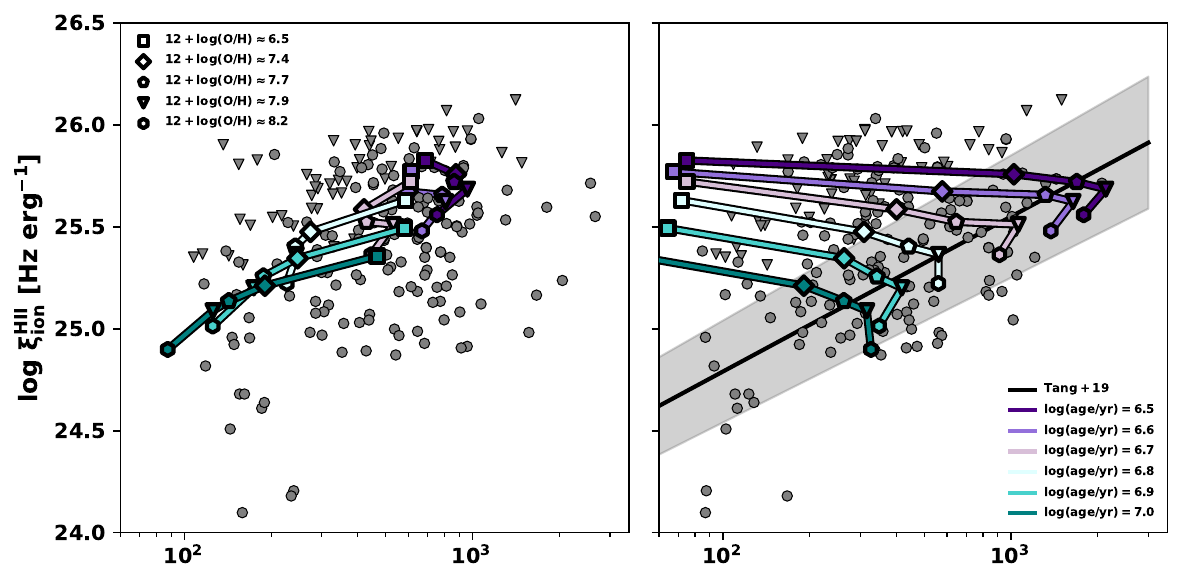}
     \includegraphics[width=\textwidth]{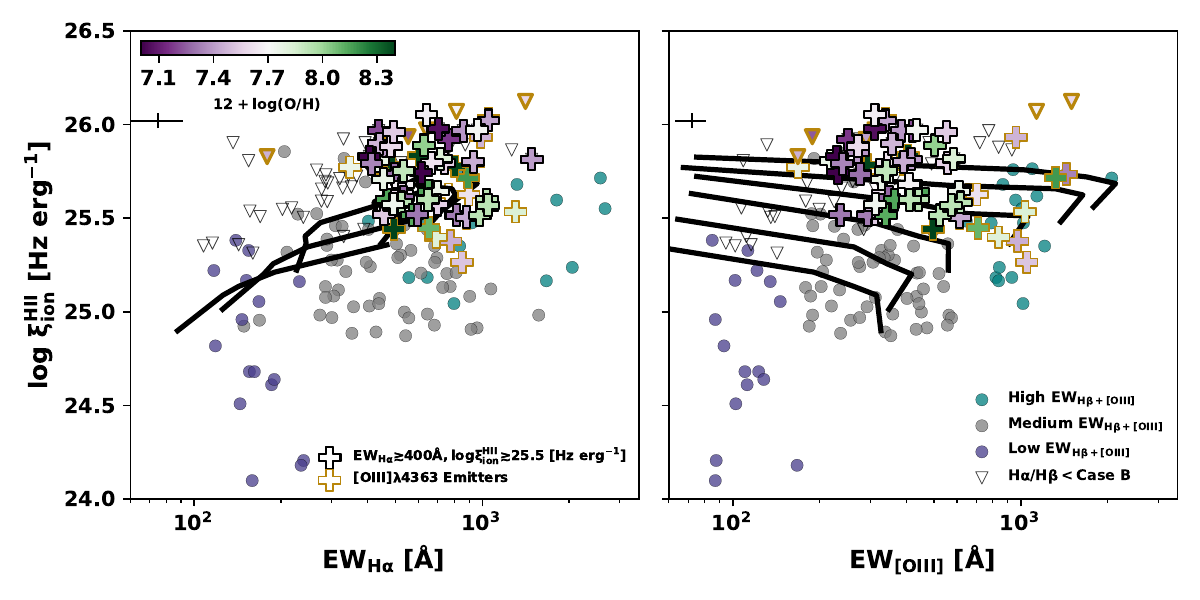}
     \caption{\textit{Top Left}: Change in age of starburst at fixed log(U)=-2 with \texttt{BPASS+CLOUDY} models including dust in H$\alpha$~EW space; we indicate the respective model metallicity changes as well. As assumed, metallicity is minor compared to the effect of log(U) between H$\alpha$~EW and $\rm \xi^{HII}_{ion}$. \textit{Top Right}: The same \texttt{BPASS+CLOUDY} models but in [OIII]EW space. We see the \texttt{BPASS+CLOUDY} predict a non-linear leftward offset at fixed $\rm \xi^{HII}_{ion}$ when metallicities are lower than $\rm 12+log(O/H) \lesssim 7.70$ ($\rm 10\% Z_{\cdot}$), similar to our observed leftward offset ($\sim 1$dex) in our medium EW sample exhibiting comparable $\rm \log \xi^{HII}_{ion}$ values (Figure \ref{Ionization Efficiency vs EWs}). Between model metallicities of $\rm 7.70 \lesssim 12+log(O/H) \lesssim 8.20$~(i.e., the dynamic range in which [OIII]'s sensitivity to metallicity variations is reduced), increasing the time from the star-formation event reproduces the derived relation from \citetalias{Tang_2019} (shaded region).~\textit{Bottom Left}: Same as above but medium EW galaxies with $\rm \log \xi^{HII}_{ion} \geq 25.5~[Hz~erg^{-1}]$~and [OIII]$\lambda 4363$~emitters are color-coded by metallicity (Section \ref{Metallicity Derivations and Photoionization Modelining}). The \texttt{BPASS+CLOUDY} models predict a diminishing metallicity effect at high H$\alpha$~EWs, aligning with the observed scatter in metallicity measurements.~\textit{Bottom Right}: The efficient ionizers from the medium EW sample and [OIII]$\lambda 4363$~emitters in [OIII] space. A metallicity gradient is observed to lower [OIII] EWs and higher $\rm \log \xi^{HII}_{ion}$ values, aligning with the predicted trends of the \texttt{BPASS+CLOUDY} models. This gradient occurs since galaxies with metallicities less than 12+log(O/H)$\lesssim 7.70~(10\%Z_{\odot})$ are disproportionally offset to lower [OIII] EWs relative to more enriched counterparts with similar ionization efficiencies. }
     \label{age metallicity model}
\end{figure*}

We first investigate if variations in the ionization parameter (U) could lead to the systematic offset we observe. We fix our \texttt{BPASS+CLOUDY} models at $\approx 2$~Myr post star-forming event, finding our models predict increased [OIII]$\lambda 5007$~EW leftward-offset at fixed $\rm \xi^{HII}_{ion}$ when large variations in $\rm \log(U)$ are present ($\rm -1 \gtrsim log(U) \gtrsim -3$). However, we observe the \texttt{BPASS+CLOUDY} models reproduce our H$\alpha$~EW distribution and scatter when $\rm \log(U)$ is between $\rm \approx -1.5~and~-2$, as expected. In general, to account for the full [OIII]$\lambda 5007$ EW offset, $\rm \log(U)$~values $\gtrsim -1$~would be required, well beyond the reported values of SFGs in the high-$z$~Universe \citep[e.g.,][]{Reddy_2023} and what is typically predicted \citep{Blanc_2015}. We therefore investigate the offset induced with age and metallicity by fixing our \texttt{BPASS+CLOUDY} models to $\rm \log(U) = -2.0$, varying ages between $\approx 2-10$~Myrs, and tracking \texttt{BPASS} metallicities in our parameter spaces. We see in the left panel of Figure \ref{age metallicity model} that the observed relation between H$\alpha$ EW and $\rm \xi^{HII}_{ion}$ is largely set by the age of the star-formation event, with metallicity accounting for less scatter compared to $\rm \log(U)$. While in the right panel of Figure \ref{age metallicity model}, the observed \textit{scatter} between [OIII]$\lambda 5007$ EW and $\rm \xi^{HII}_{ion}$ cannot be fully explained by starburst age, but the \textit{relation} itself likewise is. In fact, at fixed metallicity, increasing the time since the star-forming event traces the relation from \citetalias{Tang_2019}, while simultaneously, the \texttt{BPASS+CLOUDY} models demonstrate a $\sim 0.7$dex difference in [OIII]$\lambda 5007$ EWs between $\rm 6.5 \lesssim 12+\log(O/H) \lesssim 7.7$. In other words, extremely low-metallicity models indicate the largest directional offset between [OIII]$\lambda 5007$ EW and $\rm \xi^{HII}_{ion}$ relative to $z \sim 2$ relations. 

It is advantageous to then focus on the medium EW sample overlap in H$\alpha$~EW space. We therefore select all galaxies from our medium EW sample with H$\alpha$~EWs $\geq 400$\AA~and $\rm \log \xi^{HII}_{ion} \geq 25.5$~[Hz erg$^{-1}$]. We present in the lower panel of Figure \ref{age metallicity model} our [OIII]$\lambda 4363$ sample and medium EW subsample overlaid on the \texttt{BPASS+CLOUDY} models from the upper panel. We color-code these galaxies by metallicity, indicating $\rm \sim 10\% Z_{\odot}$ as the turnover color. We require a $\rm 2\sigma$~[OII]$\lambda 3727,29$ detection to derive metallicities using~$\rm \hat{R}$, else we employ the R3 calibration from \cite{Sanders_2023} (instead of reporting metallicity upper limits). Both $\rm \hat{R}$~and R3 are double branched, and we do not significantly detect [NII]$\lambda 6584$~or [SII]$\lambda6717,31$~in this population to break the degeneracies. Instead, we use our [OII]$\lambda 3727,29$~measurements and $2\sigma$ upper limits derived directly from the noise spectrum following \cite{Cameron_2023} as the degeneracy breaker for the calibrations.

As predicted from the \texttt{BPASS+CLOUDY} models, we observe a general offset between derived metallicities with increasing $\rm \xi^{HII}_{ion}$ and decreasing EW$\rm_{[OIII]}$~that is not necessarily observed at fixed EW$\rm_{H\alpha}$. $71\%$~$\rm (22/31)$ of our [OIII]$\lambda 4363$ population is less than $\rm \sim 10\% Z_{\odot}~(median~12+log(O/H) = 7.54)$, coinciding with the greater dynamic range of our \texttt{BPASS+CLOUDY} models; likewise, there is the tendency of more metal-poor galaxies in our medium EW subsample to possess lower EW$\rm_{[OIII]}$, aligning with the non-linear metallicity effects on EW$\rm_{[OIII]}$. 

We do observe a few enriched systems ($\rm 12+log(O/H) \gtrsim 8.0$) above $z \sim 2$~relations at moderate [OIII] EWs ($\rm \sim 400-600\AA$), however. These metallicities are near where our models predict EW$\rm_{[OIII]}$~peaks at fixed $\rm \xi^{HII}_{ion}$, but \cite{Laseter_2023} and \cite{Sanders_2023} demonstrated elevated excitation ratios at these metallicities in the high-$z$~Universe. It is therefore expected for more chemically evolved systems to scatter upward in the presence of harder ionization fronts and starbursting episodes, especially as $\rm \xi^{HII}_{ion}$~is directly dependent on ionizing photon production. Extremely metal-poor AGNs are still an option even though we have not detected any broad or high ionization lines as an AGN presence would result in erroneous $\rm T_e$~values based on our current derivations \citep{Dors_2021}. Regardless, these outliers underscore the overlay between extremely low metallicities and starbursting episodes in more evolved systems.

In fact, our extremely metal-poor galaxies deviate from the notion that the most intense (and efficient) radiation fields are exclusively found at the most extreme [OIII] EWs ($\rm \gtrsim 600\AA$). It is not that extreme [OIII] EWs lack young stellar populations, but that the effect of metallicity cannot be ignored when selecting with [OIII] in the high-$z$~Universe. The high H$\alpha$~EWs we measure for our extremely metal-poor galaxies reflect sSFRs associated with the lowest mass and youngest galaxies experiencing starbursting episodes above the star-forming main sequence, aligning with our measured ionization efficiencies. However, low-metallicities have the effect of increasing H$\alpha$~EW, so our interpretations are ambiguous so far, requiring further discussion of $\rm \xi^{HII}_{ion}$.

%%%%%%%%%%%%%%%%%%%%%%%%%%%%%%%%%%%%%%%%%%%%%%%%%
%%%%%%%%%%%%%%%%%%%%%%%%%%%%%%%%%%%%%%%%%%%%%%%%%
%%%%%%%%%%%%%%%%%%%%%%%%%%%%%%%%%%%%%%%%%%%%%%%%%%%%%%%%%%%%%%%%%%%%%%
\section{\textbf{Discussion}} \label{Discussion}
%%%%%%%%%%%%%%%%%%%%%%%%%%%%%%%%%%%%%%%%%%%%%%%%%%%%%%%%%%%%%%%%%%%%%%

%%%%%%%%%%%%%%%%%%%%%%%%%%%%%%%%%%%%%%%%%%%%%%%%%%%%%%%%%%%%%%%%%%%%%%
\subsection{\rm \textbf{Nature and Selection of Extremely Metal-Poor Efficient Ionizers}} \label{Selection of EELGs/Efficient Ionizers}
%%%%%%%%%%%%%%%%%%%%%%%%%%%%%%%%%%%%%%%%%%%%%%%%%%%%%%%%%%%%%%%%%%%%%%

We have spectroscopically demonstrated extremely metal-poor efficient ionizers are common above $z \gtrsim 2$, and that the dissimilar trends between Balmer and [OIII] EWs with $\rm \xi^{HII}_{ion}$ in the high-$z$ Universe are caused primarily by oxygen abundances lower than $\rm 12+log(O/H) \approx 7.7$ ($\rm10\%Z_{\odot}$). These galaxies are comparatively rare in pre-JWST $z \sim 2$ and $z \sim 7$ samples as it is above $\rm 12+\log(O/H) \approx 7.7$ where oxygen has a weak dependence on metallicity yet high [OIII]/H$\beta$ ratios \citep[e.g., R3 and R23,][]{Maiolino_2019}, especially as the dynamic range of [OIII]'s sensitivity is reduced in the high-z Universe \citep{Sanders_2023, Laseter_2023}. Metallicity effects were therefore diminished and the luminosity-weighted ages setting Balmer and [OIII] EW relations with $\rm \xi^{HII}_{ion}$ were more apparent, though \citetalias{Tang_2019} did predict redshift evolution in [OIII]. 

\begin{figure*}[hbt!]
    \centering
    \includegraphics[width=\textwidth]{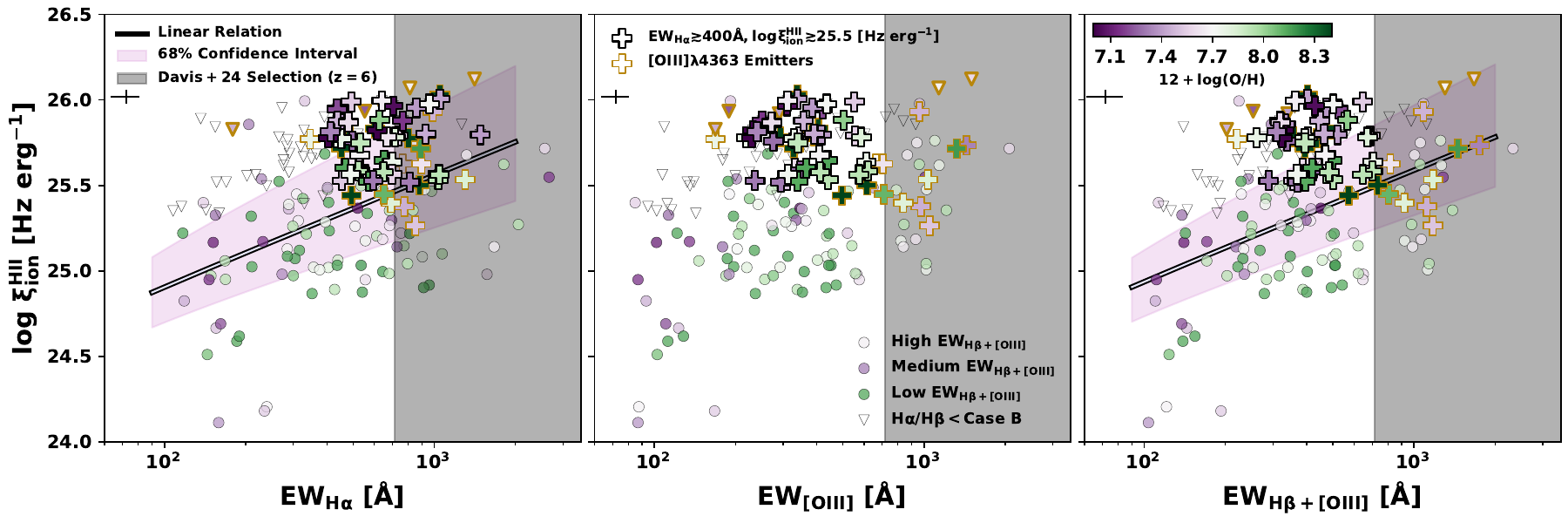}
    \caption{Correlation between $\rm \xi^{HII}_{ion}$ and H$\alpha$~EW (\textit{Left}), [OIII] EW (\textit{Center}), and H$\beta$+[OIII] EW (\textit{Right}) with data color-coded to their $\rm T_{e}$-derived or $\rm \hat{R}$-derived metallicities. Equation \ref{ha equation} and our H$\beta$+[OIII]~derived linear relation (see Section \ref{The Use of Nebular EWs as Proxy for Ion Eff}) are overlaid with $68\%$ confidence intervals. The grey-shaded region represents the selection function of \citetalias{Davis_2023} at $z = 6$. It is apparent EW$\rm_{H\alpha}$ is a more direct selection of extremely metal-poor ionizers compared to a selection involving [OIII]. The H$\beta$+[OIII] EELG selection function of \citetalias{Davis_2023}, which is reasonable based on prior [OIII]-$\rm \xi^{HII}_{ion}$~scaling relations, disproportionally excludes these galaxies compared to a selection on H$\alpha$~alone. This effect is also seen in our derived relations considering we underestimate $\rm \xi^{HII}_{ion}$~for the extremely metal-poor population when using H$\beta$+[OIII] yet align with the H$\alpha$~relation found in \citetalias{Tang_2019} that better predicts $\rm \xi^{HII}_{ion}$~within our confidence region. The physical limitations of $\rm \xi^{HII}_{ion}$~combined with EWs $\rm > 1000\AA-1500\AA$~in both H$\alpha$~and~H$\beta$+[OIII] requiring more careful treatment (e.g., dominant nebular contributions \citep{Katz_2024} and AGNs \citepalias{Davis_2023})~is likely decreasing the apparent metallicity trends at EW$\rm_{H\beta+[OIII]}~\gtrsim 1000$~in our sample. Within the general high-$z$~EW distribution though, a sole selection on H$\beta$+[OIII] EW will bias to more chemically evolved starbursting systems at fixed $\rm \xi^{HII}_{ion}$.}
    \label{Missed Subsample}
\end{figure*}

Specifically, $\rm \xi^{HII}_{ion}$ does not have a strong dependence on metallicity relative to EW([OIII]) \citep{Katz_2024}, resulting in a flattening within this parameter space at low metallicities. In a compounding effect, decreasing metallicity disproportionally drives [OIII] EW below or near typical selection limits of EELGs, thus limiting the identification of extremely metal-poor ionizers. For example, \cite{Davis_2023} \citepalias{Davis_2023}, selected 1165 EELGs in CEERS by imposing an observed frame $\rm EW \geq 5000\AA$ cutoff~in $\rm H\beta + [OIII]\lambda 5007$ and $\rm H\alpha$ (i.e., rest-frame EW $> \rm 1000\AA, 714\AA$ for $z = 4,6$). \citetalias{Davis_2023} found their sample density was high near this detection limit, suggesting a continuous distribution to lower EWs. We present in Figure \ref{Missed Subsample} the [OIII], H$\alpha$, and H$\beta$+[OIII] EW distribution with $\rm \xi^{HII}_{ion}$~color coded by metallicity, including the \citetalias{Davis_2023} selection limit at $z=6$.

We observe a continuous distribution to lower H$\alpha$ and $\rm H\beta+[OIII]$~EWs, but with the key detail that high $\rm \xi^{HII}_{ion}$ values are maintained down to EW$\rm_{H\beta+[OIII]} \approx 300\AA-400\AA$. Regarding our medium EW subsample, $\rm 98\% (51/52)$ of these galaxies overlap with the H$\alpha$~EW distribution of our high EW sample, while $\rm 0\%$ overlap with the H$\beta$+[OIII] distribution. Therefore, in addition to a continuous distribution to lower H$\beta$+[OIII] EWs at fixed $\rm \xi^{HII}_{ion}$, we find that H$\beta$+[OIII] EW is not a one-to-one selection with H$\alpha$. 

The effectiveness of H$\alpha$~EW selecting extremely metal-poor ionizers reflects the underlying trends of H$\alpha$~EW with sSFR \citep[e.g.,][]{Marmol_2016}. However, high H$\alpha$~EWs do not exclusively mean higher sSFRs considering the effect of line-blanketing driving H$\alpha$~emission at low metallicities \citep{Dreizler_1993, Grafener_2002}. For example, \cite{Endsley_2023b} followed up \citetalias{Endsley_2023} with NIRCam imaging from JADES (DR1), finding dissimilar H$\beta$+[OIII] distributions between their faint ($\rm M_{UV} \approx -17.5$) and bright ($\rm M_{UV} \approx -20$) samples yet nearly identical H$\alpha$ distributions, similar to our findings. \cite{Endsley_2023b} derived a $\sim 1$dex decrease in metallicity (\texttt{Beagle}-derived) between their bright and faint samples, with the peak of the faint distribution being at 12+log(O/H) = 7.40 ($\sim 5\%Z_{\odot}$) and the bright at 12+log(O/H) = 8.06 ($\sim 25\%Z_{\odot}$). Such a decrease in metallicity results in a $50\%$~ increase in H$\alpha$~EW according to \cite{Endsley_2023b} using the \cite{Gutkin_2016} models. As such, \cite{Endsley_2023b} proposed low metallicities in the faint population, but with the addition of a relative downturn in star formation to not increase H$\alpha$ EW relative to their UV bright population, reflecting bursty star formation. Considering our medium EW sample extends down to $\rm M_{UV} \approx -17$ while having weakened H$\beta$+[OIII] EWs along the \texttt{BPASS+CLOUDY} model age traces in H$\alpha$ (Figure \ref{age metallicity model}), chemically poor galaxies in a relative downturn in star-formation are naturally in our sample. 

However, our \texttt{BPASS+CLOUDY} models suggest a diminishing effect of metallicity in driving H$\alpha$~EWs with decreasing starburst age, such that at more extreme starbursts there is not a drastic H$\alpha$~EW increase. The median H$\alpha$ EWs of our high EW and medium EW subsample are $\rm 896\AA~and~628\AA$~respectively, well within the range where metallicity effects are reduced. Moreover, it is below 12+log(O/H)$\approx 7.0~(2\%Z_{\odot})$ where the largest increase of H$\alpha$~occurs, but given the nebular metallicity “floor” \citep{Kobayashi_2024} and our derived metallicities, it is unlikely that we have a large sample of sub 12+log(O/H)=7.0 galaxies inflating H$\alpha$~EWs to the extremes we measure. We also measure $\rm \log \xi^{HII}_{ion}$~values greater than $\rm25.5~[Hz~erg^{-1}]$~ down to EW$\rm_{[OIII]} \approx 300\AA$. Recalling the specific definitions of ionization efficiency outlined in Section \ref{The Ionizing Photon Production Efficiency}, \cite{Katz_2024} demonstrated across various SPS models that the observed ionization efficiency ($\rm \xi^{HII}_{ion}$) and the ionization efficiency associated with the intrinsic stellar population ($\rm \xi^{*}_{ion}$) agree below $\rm \log \xi^{HII}_{ion} = 25.5~[Hz~erg^{-1}]$, whereas above $\rm \xi^{HII}_{ion}$~deviates low compared to $\rm \xi^{*}_{ion}$. Therefore, our high $\rm \xi^{HII}_{ion}$~values are expected to be associated with even higher intrinsic ionization efficiencies from young stellar populations. \cite{Katz_2024} considered ages up to $\rm 20$~Myr assuming an instantaneous burst, though at these extremely low metallicities, the inferred star-formation histories are likely younger. For example, the majority of the bright population from \cite{Endsley_2023b} is distributed between $\rm \log \xi^{HII}_{ion} \approx 25.5-25.6~[Hz~erg^{-1}]$ with an average positive SFR$\rm_{3Myr}$/SFR$\rm_{50Myr}$ ($\sim 1.5$). This $\rm \log \xi^{HII}_{ion}$~median is comparable to the median of our medium EW subsample ($\rm \log \xi^{HII}_{ion} = 25.7~[Hz~erg^{-1}]$), but in contrast, the average metallicity of their bright population is 12+log(O/H)$\rm \approx 8.0-8.1~(20\%-25\%Z_{\odot})$ compared to 12+log(O/H)$\rm \approx 7.5-7.6~(7\%-8\%Z_{\odot})$~for our specific population, suggesting more pristine ISM conditions. We argue then that in addition to the findings of the more general population studies of \citetalias{Endsley_2023} and \cite{Endsley_2023b}, there exists extremely metal-poor ionizers exhibiting large H$\alpha$~EWs ($\rm \gtrsim 600\AA$) that are indeed starbursting galaxies rather than ``artificially" boosted metal-poor galaxies in a relative starbursting lull. We note that our emission line selection (Section \ref{The Final Samples}) preferentially allows us to find these low-metallicity galaxies as opposed to systems with SF downturns, as found in \citetalias{Endsley_2023} and \cite{Endsley_2023b}.

\begin{figure*}
    \centering
    \includegraphics[width=\textwidth]{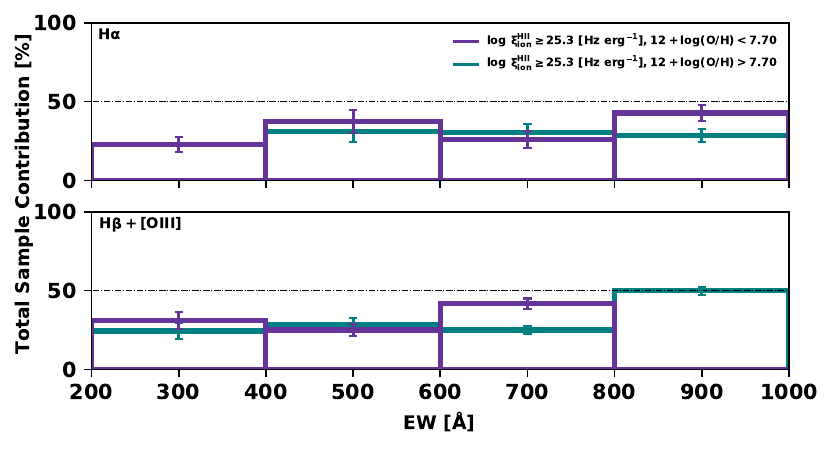}
    \caption{\textit{Top}: EW$\rm_{H\alpha}$ and the contributing percentage of extremely metal-poor (12+log(O/H$< 7.7$; purple) and enriched (12+log(O/H$> 7.7$; green) efficient ionizers ($\rm \log \xi^{HII}_{ion} \gtrsim 25.3~[Hz~erg^{-1}]$). Metallicity subsamples contribute comparable amounts to the H$\alpha$~EW and $\rm \xi^{HII}_{ion}$~distribution, reflecting the dominant dependency of H$\alpha$~EW and $\rm \xi^{HII}_{ion}$~on ionizing photon production compared to metallicity. \textit{Bottom:} The contributing fraction of efficient ionizers with respect to EW$\rm_{H\beta+[OIII]}$. The contributing percentages of our respective metallicity samples deviate with EW$\rm_{H\beta+[OIII]}$, largely an effect from extremely metal-poor galaxies offsetting disproportionally left from more chemically enriched systems with similar $\rm \xi^{HII}_{ion}$~values.}
    \label{Sample_Fraction}
\end{figure*}

Regardless of how one quantifies burstiness, we demonstrate that customary $\rm H\beta + [OIII]$~EW selections (e.g., EW(H$\beta$+[OIII])$> 700$\AA) favor more chemically evolved SFGs around $7.8 \lesssim $12+log(O/H)$\lesssim 8.3$~($\gtrsim 20\%Z_{\odot}$) compared to a selection on Balmer EW alone. We present in Figure \ref{Sample_Fraction} the fraction of extremely metal-poor (12+log(O/H)$<7.7$) and enriched (12+log(O/H)$>7.7$) efficient ionizers ($\rm \log \xi^{HII}_{ion} \geq 25.3~[Hz~erg^{-1}]$) relative to our total sample in $\rm 200\AA$~EW bins. We choose $\rm \log \xi^{HII}_{ion} \geq 25.3~[Hz~erg^{-1}]$~as this exceeds the so-called ``canonical value" ($\rm \log \xi^{HII}_{ion} = 25.2-25.3~[Hz~erg^{-1}]$) \citep{Robertson_2013}. As expected, metallicity subsamples contribute comparable amounts to the H$\alpha$~EW and $\rm \xi^{HII}_{ion}$~distribution, reflecting the dominant dependency of H$\alpha$~EW and $\rm \xi^{HII}_{ion}$~on ionizing photon production compared to metallicity. The near constant $50\%$~contribution of efficient ionizers reflects the intrinsic scatter of our sample being nearly symmetric around the relation from \citetalias{Tang_2019}. More interestingly, there are dissimilar trends in the contribution of extremely metal-poor and enriched systems dependent upon EW$\rm_{H\beta+[OIII]}$. This effect is largely from the observed metallicity gradient in Figure \ref{Missed Subsample} where extremely metal-poor galaxies exhibit H$\beta$+[OIII] EW distributions of values similar to more evolved systems that have begun to lose their most massive stars, resulting in a separation with $\rm \xi^{HII}_{ion}$. These extremely metal-poor galaxies then reside in EW$\rm_{H\beta+[OIII]}$~ distributions of more evolved systems that have begun to lose their most massive stars, resulting in a separation with $\rm \xi^{HII}_{ion}$. The effect is then at typical EW distributions of the high-$z$~Universe \citep[e.g.,][]{Boyett_2024, Heintz_2024, Roberts_2024}, a sole selection on H$\beta$+[OIII] EW will bias to more 1) chemically enriched systems experiencing similar ionizing fronts and 2) evolved starbursting systems. 

However, the EW$\rm_{H\beta+[OIII]}$~metallicity gradient observed in our medium EW subsample is not necessarily apparent in our high EW sample, and the contribution of extremely metal-poor efficient ionizers begins to increase above EW$\rm_{H\beta+[OIII]} \approx 1000\AA$. This is primarily due to the limited sample above EW$\rm_{H\beta+[OIII]} \approx 1000\AA$ \citep[e.g.,][]{Boyett_2024}, but in general, such high EWs are near the extremes of models with younger stellar populations \citep[e.g.,][]{Wilkins_2020}. Rest-frame H$\beta$+[OIII] EWs from $\rm 1000\AA-5000\AA$ have been measured with JWST, but they are a small percentage of the EELG distribution \citep{Boyett_2024}. Additionally, the contributions of AGN and nebular continuum are beginning to be explored in these populations, for example, some higher EWs correspond to more compact emission regions \citepalias{Davis_2023} and the EW distribution of nebular-dominated galaxies exceeds the general population of EELGs \citep{Katz_2024}. Additionally, the physical limit of $\rm \xi^{HII}_{ion}$~ has the effect of flattening the observed EW-$\rm \xi^{HII}_{ion}$~trend and creating a plateau at extremely high EWs. As such, we do not expect our observed trends to necessarily hold past EW$\rm_{H\beta+[OIII]} \sim 1500\AA$ as metallicity effects are diminished compared to possible non-stellar ionizing sources and extreme nebular temperatures and densities \citep{Katz_2024}. Regardless, extremely metal-poor SFGs within more typical high-$z$~EW distributions have non-negligible metallicity effects when measuring EW$\rm_{H\beta+[OIII]}$. 

Because [OIII]-deficient, ionizing systems are comparatively missed under normal H$\beta$+[OIII] selections, it is uncertain whether H$\beta$+[OIII] selected EELGs are representative of the entire population of extremely metal-poor efficient ionizers. Systems with high sSFRs at the lowest metallicities are key in characterizing high-$z$ initial mass functions and star-formation histories; key frameworks that have already been invoked in more significant manners in the high-$z$ Universe, e.g., more top-heavy \citep{Cameron_2024} and bursts \citep{Pallottini_2023, Sun_2023, Sun_2023b, Topping_2024b}. These systems are also interesting for investigating the onset and evolution of the mass-metallicity relation and fundamental-metallicity relation \citep[e.g.,][]{Curti_2023, Curti_2023b} given the pristine gas likely in these galaxies. A deeper examination of the relative abundance ratios for our most metal-poor efficient ionizers is of interest as well considering the various proposed enrichment scenarios of peculiar abundance ratios found at high-$z$ \citep[e.g.,][]{Cameron_2023b, Kobayashi_2024}. Fortunately, the most metal-poor SFGs can be identified by a ratio of H$\alpha$ to H$\beta$+[OIII] photometrically. Still, a deeper examination at fainter magnitudes is required considering that \cite{Simmonds_2024} find a minor increase in $\rm \xi^{HII}_{ion}$ for fainter galaxies with burstier SFHs in JEMS and JADES NIRCam imaging. Additionally, \cite{Endsley_2023b} found two-component and continuous star-formation histories differ drastically in \texttt{Beagle} EoR predictions, with continuous models systemically increasing $\rm f_{esc}$ to fainter $\rm M_{UV}$ SFHs and two-component models assuming small $\rm f_{esc}$ regardless of $\rm M_{UV}$, thus forcing a downturn in star-formation within their faint populations. \cite{Choustikov_2024} demonstrated, however, that a single parameter cannot predict $\rm f_{esc}$~without large scatter, such that $\rm f_{esc}$ requires several observable quantities (eight in their model). In this context and that of \cite{Simmonds_2024}~findings, exploring highly efficient ionizers in extremely metal-poor environments with the theoretical bases of \cite{Choustikov_2024} and \cite{Katz_2024} could better characterize the role of the faint end of the UV luminosity function in Reionization. Likewise, the escape of ionizing photons is not instantaneous with their production \citep{Choustikov_2024, Saxena_2024}. It is of interest then to examine the emergence of escape channels in the most extremely metal-poor efficient ionizers, i.e., how $\rm \xi^{HII}_{ion}$~values corrected for $\rm f_{esc}$~scale with more general properties (e.g., surface SFR and gas density) in the presence of young stellar populations.

%%%%%%%%%%%%%%%%%%%%%%%%%%%%%%%%%%%%%%%%%%%%%%%%%%%%%%%%%%%%%%%%%%%%%%
\subsection{\rm \textbf{The Use of Nebular EWs as Proxy for $\rm \xi^{HII}_{ion}$}} \label{The Use of Nebular EWs as Proxy for Ion Eff}
%%%%%%%%%%%%%%%%%%%%%%%%%%%%%%%%%%%%%%%%%%%%%%%%%%%%%%%%%%%%%%%%%%%%%%

The discussion from the preceding section indicates EW(H$\alpha$) is a strong indicator of $\rm \xi^{HII}_{ion}$. We therefore provide an updated relation between EW(H$\alpha$) and $\rm \xi^{HII}_{ion}$ by fitting an error-weighted linear function for $10,000$~iterations:
\begin{equation} \label{ha equation}
    \begin{aligned}
        \rm \log \xi^{HII}_{ion}~[Hz~erg^{-1}] = (0.63 \pm 0.14)\times
        log(EW_{H\alpha}~\AA)\\
        + (23.64 \pm 0.35).
    \end{aligned}
\end{equation}
We present our derived relation with $1\sigma (68\%)$ confidence regions in the left panel of Figure \ref{Missed Subsample}. The slope of Equation \ref{ha equation} is $37\%$~ shallower than \citetalias{Tang_2019} and has a $2.7\%$~higher intercept. These differences likely reflect the methodology in calculating $\rm L_{UV}$. In particular, \citetalias{Tang_2019} inferred $\rm L_{UV}$ from their best-fitting \texttt{Beagle} models using a $100$\AA~bin centered at $1500$\AA, and thus assumed a photoionization model and set of parameters. In comparison, we do not employ stellar population synthesis models in our continuum fitting and we extend our UV range to longer wavelengths (Section \ref{UV Luminosities}), thus small $\rm \xi^{HII}_{ion}$ differences at fixed H$\alpha$~EW are expected. However, Equation \ref{ha equation} introduces large uncertainties in $\rm \xi^{HII}_{ion}$ as we have significant scatter in our $\rm \xi^{HII}_{ion}$ relative to \citetalias{Tang_2019}; \citetalias{Tang_2019} measures $\sigma = 0.20$ while we measure a standard deviation of $\sigma = 0.35$, a $\sim 43\%$ increase. This is not surprising as $\rm \xi^{HII}_{ion}$ is dependent on the ionizing photon production and underlining UV continuum. High-$z$ star-formation histories, for example, are diverse in duration and intensity \citep{Looser_2023, Looser_2024}, from which heterogeneity in metal enrichment from type-I and type-II supernovae leading to changes in the ionization output of high-mass stars is expected. In addition, variations in the slope of the initial mass function (IMF) and nebular continuum contribution can significantly affect the UV at $\rm1500\AA$, further increasing the observed scatter. Different dust conditions, such as dust exposure to harder radiation fronts absent from standard dust laws, thus leading to an underestimation of the UV luminosity when employing Balmer decrements \citep{Markov_2023, Sanders_2024}, could also play a role, amongst variations in electron density, non-stellar ionizing sources, or more exotic stellar populations \citep[for a deeper discussion see][]{Katz_2024}. Nonetheless, we largely agree with the H$\alpha$-$\rm \xi^{HII}_{ion}$ relation from \citetalias{Tang_2019}, albeit with greater variation.

In comparison, we find it challenging to describe the relation between [OIII] and $\rm \xi^{HII}_{ion}$ as a simple scaling law. The conflation of metallicity with ionization and age leads to a systematic underprediction of $\rm \xi^{HII}_{ion}$ in extremely metal-poor ionizers.
We perform an error-weighted linear function for the quantities $\rm H\beta + [OIII]$~ EW and $\rm \xi^{HII}_{ion}$, finding that assuming a linear relationship results in a $\rm 0.5$dex underestimation, though the diversity of our H$\alpha$ sample makes this relation within error of our H$\alpha$~EW relation (Equation \ref{ha equation}). As we mention in Section \ref{Selection of EELGs/Efficient Ionizers} and demonstrate in Figure \ref{Missed Subsample}, however, these representative samples predominately overlap in H$\alpha$~EW space, suggesting similar star-formation histories and that metallicity degeneracies can be alleviated with sSFR measurements. We, therefore, caution that linear [OIII] EW-$\rm \xi^{HII}_{ion}$~relations may not be applicable in the high-$z$~Universe. 

%%%%%%%%%%%%%%%%%%%%%%%%%%%%%%%%%%%%%%%%%%%%%%%%%%%%%%%%%%%%%%%%%%%%%%
\section{\textbf{Summary}} \label{Summary}
%%%%%%%%%%%%%%%%%%%%%%%%%%%%%%%%%%%%%%%%%%%%%%%%%%%%%%%%%%%%%%%%%%%%%%

We investigate a population calling into question the ubiquity of young galaxies possessing high H$\beta~+$~[OIII] EWs with JADES NIRSpec spectroscopy; we summarize our methods and findings as follows: 
 
\begin{enumerate}

    \item We determine how the distribution of the ionizing photon efficiency ($\rm \xi^{HII}_{ion}$) scales with H$\alpha$, [OIII]$\lambda 5007$, and H$\beta$ + [OIII]$\lambda 5007$ EWs for $z \gtrsim 2$~(Figure \ref{Ionization Efficiency vs EWs}). We find dissimilar trends between $\rm \xi^{HII}_{ion}$ and the different line EWs, suggesting a physically motivated difference in the EW distribution of our sample. When we divide our sample into low, medium, and high H$\beta$ + [OIII] EW samples, we find $\rm 37\%$ of our ``medium" EW sample has elevated H$\alpha$ EWs similar to our high EW sample. This finding implies that weakened observed H$\beta$ + [OIII]$\lambda 5007$ EWs are driven by suppressed [OIII] emission compared to H$\beta$, indicative of low gas-phase metallicities. The combination of our H$\alpha$~EW and $\rm log\xi^{HII}_{ion}$~measurements $\rm \gtrsim 25.5~[Hz~erg^{-1}]$~also suggests these faint, extremely metal-poor galaxies are indeed experiencing intense and recent star formation episodes, aligning with the notion of stochastic star formation in the context of the larger population studies of \cite{Endsley_2023, Endsley_2023b}.
    
    \item  We combine \texttt{BPASS+CLOUDY} models with $\sim 31$~[OIII]$\lambda 4363$~emitters and $\rm \hat{R}$-derived metallicities, finding metallicities less than 12+log(O/H)$\rm \approx 7.70~(10\%Z_{\odot})$~significantly reduce H$\beta$+[OIII] EWs without impacting $\rm \xi^{HII}_{ion}$. This non-linear effect is observed in a clear metallicity gradient to lower H$\beta$+[OIII] EWs and higher $\rm \xi^{HII}_{ion}$~(Figure \ref{age metallicity model}). We further demonstrate these extremely metal-poor ionizers are common above $z \sim 2$, yet comparatively rare in pre-JWST $z \lesssim 2$ and $z \sim 7$ studies. It is above $\rm 12+\log(O/H) \approx 7.7$ where oxygen has a weak dependence on metallicity yet high [OIII]$5007$/H$\beta$ ratios, and so intense [OIII] emitters were more likely to be identified, but in return, they were likely to be more chemically enriched ($\rm 7.7 \lesssim 12+\log(O/H) \lesssim 8.3$), especially as the dynamic range of [OIII]'s metallicity sensitivity is reduced in the high-z Universe. Metallicity effects were diminished and the luminosity-weighted ages setting Balmer and [OIII] EW relations with $\rm \xi^{HII}_{ion}$ were more apparent.
    
    \item We demonstrate extremely metal-poor efficient SFGs are excluded under reasonable extreme emission line galaxy H$\beta$ + [OIII] selections, e.g., EW(H$\beta$+[OIII])$ > 700$\AA~(Figure \ref{Sample_Fraction}). A sole selection of H$\beta$+[OIII] has a bias towards 1) enriched systems with similar extreme radiation fields and 2) more evolved star forming systems as the effect of extremely low metallicities at fixed $\rm \xi^{HII}_{ion}$~is non-linear. However, these systems can be identified through their H$\alpha$~equivalent widths.

    \item We find our sample generally follows the H$\alpha$ EW and $\rm \xi^{HII}_{ion}$ relation from \cite{Tang_2019}, further suggesting that high-$z$ systems should be identified based on their H$\alpha$ emission, i.e., their respective location along the star-formation main-sequence (Figure \ref{Missed Subsample}). However, we find significant scatter in our data, likely reflecting different ionization, age, dust, and nebular conditions. We find there is added degeneracy in interpreting H$\beta$+[OIII] EW as a tracer of ionization as [OIII] EW is greatly reduced below 12+log(O/H)$\rm \approx 7.70~(10\%Z_{\odot})$~ leading to $\sim 1$dex scatter above $\rm H\beta + [OIII] \approx 300\AA$. We further demonstrate the ionization efficiencies of our metal-poor galaxies are under-predicted when using $\rm H\beta + [OIII]$~scaling relations, and thus we argue against [OIII] as a sole tracer of high-$z$ efficient ionizing systems.  

\end{enumerate}

\section{Acknowledgments}
This material is based upon work supported by the National Science Foundation Graduate Research Fellowship under Grant No. 2137424. MVM is supported by the National Science Foundation via AAG grant 2205519, the Wisconsin Alumni Research Foundation via grant MSN251397, and NASA via STScI grant JWST-GO-4426. ECL acknowledges the support of an STFC Webb Fellowship (ST/W001438/1). AJB acknowledges funding from the “FirstGalaxies” Advanced Grant from the European Research Council (ERC) under the European Union’s Horizon 2020 research and innovation program (Grant Agreement No. 789056). ST acknowledges support from the Royal Society Research Grant G125142. YZ acknowledges support from the JWST/NIRCam contract to the University of Arizona NAS5-02015. JW gratefully acknowledges support from the Cosmic Dawn Center through the DAWN Fellowship. The Cosmic Dawn Center (DAWN) is funded by the Danish National Research Foundation under grant No. 140. S.C acknowledges support by European Union’s HE ERC Starting Grant No. 101040227 - WINGS. CS acknowledges support from the Science and Technology Facilities Council (STFC), by the ERC through Advanced Grant 695671 “QUENCH”, by the UKRI Frontier Research grant RISEandFALL.
%%%%%%%%%%%%%%%%%%%%%%%%%%%%%%%%%%%%%%%%%%%%%%%%%%%%%%%%%%%%%%%%%%%%%%
%%%%%%%%%%%%%%%%%%%%%%%%%%%%%%%%%%%%%%%%%%%%%%%%%%%%%%%%%%%%%%%%%%%%%%
%%%%%%%%%%%%%%%%%%%%%%%%%%%%%%%%%%%%%%%%%%%%%%%%%%%%%%%%%%%%%%%%%%%%%%
\bibliography{bibliography.bib}
%%%%%%%%%%%%%%%%%%%%%%%%%%%%%%%%%%%%%%%%%%%%%%%%%%%%%%%%%%%%%%%%%%%%%%
\appendix
%%%%%%%%%%%%%%%%%%%%%%%%%%%%%%%%%%%%%%%%%%%%%%%%%%%%%%%%%%%%%%%%%%%%%%
%%%%%%%%%%%%%%%%%%%%%%%%%%%%%%%%%%%%%%%%%%%%%%%%%%%%%%%%%%%%%%%%%%%%%%
\section{\rm \textbf{Case B Departure Galaxies in Our Sample}}
\label{Validity of Case B Recombination and the Effects of Dust}
%%%%%%%%%%%%%%%%%%%%%%%%%%%%%%%%%%%%%%%%%%%%%%%%%%%%%%%%%%%%%%%%%%%%%%

It is fascinating that galaxies exhibiting Balmer ratios less than the intrinsic ratio of Case B at T$=1.5\times 10^4$K parallel the systematic offset with [OIII] EW, are routinely [OIII]$\lambda 4363$~emitters, and have elevated $\rm \xi^{HII}_{ion}$~values at fixed EW. $\rm \xi^{HII}_{ion}$ is, however, highly susceptible to dust given the use of UV and H$\alpha$, and thus our consistently high $\rm \log \xi^{HII}_{ion}$ measurements above $\rm \approx 25.5~[Hz~erg^{-1}]$ are possibly biased considering we intentionally did not correct for dust for Case B departures. However, if we examine the medium and high EW Case B departure galaxies with the criteria from Section \ref{Selection of EELGs/Efficient Ionizers}, we find median H$\alpha$ EWs of $\rm \approx 787\AA$ and $\rm 713\AA$ and [OIII]EWs of $\rm \approx 852\AA$ and $\rm 306\AA$, respectively. These findings parallel our main findings, aligning with the notion that most extremely metal-poor galaxies are efficient ionizers with large H$\alpha$~EWs but disparate [OIII] EWs. We argue that this subsample is not primarily comprised of biased outliers but instead aligned with emerging pictures of high-$z$~Case B departure galaxies. For example, \cite{Scarlata_2024} identified a local object with an H$\alpha$/H$\beta$ ratio of $2.620 \pm 0.078$~using MMT spectroscopy. From their incredibly high S/N spectrum (e.g., detection of HeII and [ArIV] lines), they derive an O$^{++}$ temperature of $15380 \pm 850$K (using [OIII]$\lambda 4363$), an [OIII]/[OII] ratio of $11.80 \pm 0.74$, and an EW(H$\beta$)$= 165 \pm 14$\AA, indicating a hard ionization front and high sSFR. \cite{Scarlata_2024}, interestingly, find gas metallicities less than $10\%$ solar are required to reproduce their Balmer and oxygen line ratios, equivalent to our [OIII] deficient systems. Likewise, using the JADES public release data, \cite{McClymont_2024} found 52 (14) galaxies above $z > 2$ ($z > 5.3$) with Balmer decrements more than $1\sigma$ below Case B up to T$=2.0\times 10^4$K. \cite{McClymont_2024} found these sub-Case B galaxies tend to have higher [OIII]/[OII] ratios ($> 10$), steeper UV slopes, Ly$\alpha$ emission, and low metallicites ($\rm 12+log(O/H)<7.7$), aligning again with our extremely metal-poor efficient ionizers.

However, the physical picture proposed by the above works differs. \cite{Scarlata_2024} proposed lower H$\alpha$/H$\beta$ ratios arise from a preferential scattering of H$\alpha$ photons out of the line of sight, requiring asymmetric geometries and gas densities large enough for Balmer self-absorption and pumping to the $n=2$~state. The resultant Balmer ratios then originate from the respective optical depths of the Balmer lines. In contrast, \cite{McClymont_2024} explored Case C recombination \citep{Ferland_1999} combined with density-bounded nebulae. In this case, lower H$\alpha$/H$\beta$ ratios arise from continuum photons exciting the Lyman lines differently depending on the Lyman optical depth of the HII region, which then modifies the resultant cascade to the Balmer series. A main limitation identified by \cite{Scarlata_2024} was the resultant recombination emission luminosity diminishing due to density-bounded nebulae, so \cite{McClymont_2024} included gas turbulence indicative of the high-$z$~Universe to maintain Balmer luminosity. 

However, \cite{Yanagisawa_2024} demonstrated that both proposed scenarios require specific construction. For example, for the case where Lyman lines are optically thin, there is the formation of density-bounded nebulae from rapid quenching events ($<25$Myr) where high-density gas is briefly removed, as well as the ``fine-tuning" of the Lyman optical depths; while for the case where Balmer lines are optically thick, there needs to be gas densities high enough to populate $n=2$~level of hydrogen as well as specific line of sites and gas geometry. We cannot comment much from our current analysis about the physical picture governing Anomalous Balmer Emitters (ABEs) \citep{McClymont_2024}, but we see they are typically efficient ionizers. Considering the suggested density-bounded model forms from mini-quenched episodes, our high ionization efficiencies indicate extreme star formation is still present. The galaxies that removed their high-density gas may be back in a relative upturn of star formation, but we would then expect gas densities to approach Case B conditions unless large gas inhomogeneities are present. The proposed configuration of large Hydrogen densities with specific geometries and lines of site removes H-alpha photons to decrease the H-alpha/H-beta ratio, meaning our H-alpha derived ionization efficiencies are lower limits. However, as mentioned in Section \ref{Selection of EELGs/Efficient Ionizers}, we are near the theoretical maximum for ionization efficiency, meaning there cannot be many more ionizing photons gained back. This scenario would be fine if dust was present to decrease ionization efficiency, but ABEs have the peculiar trend of elevated H-gamma/H-beta ratios$-$the opposite direction of dust effects. A more detailed analysis and discussion are most certainly necessary but outside our current scope. Regardless, in addition to Ly$\alpha$~emission and high [OIII]$\lambda 5007$/[OII]$\lambda\lambda 3727,29$ ratios, we demonstrate ABEs possess high ionization efficiencies, have high H$\alpha$~EWs, and likely have extremely low metallicities (12+log(O/H)$< 7.7$; $\rm 10\% Z_{\odot}$) diminishing H$\beta$+[OIII] EWs and their inclusion in EELG samples.

\end{document}